\documentclass[twocolumn,showpacs,amsmath,amssymb,superscriptaddress,nofootinbib]{revtex4}

\usepackage{graphicx}
\usepackage{epsfig}
\usepackage{dcolumn}
\usepackage{bm}

\newcommand{\beq}{\begin{equation}}
\newcommand{\eeq}{\end{equation}}
\newcommand{\bea}{\begin{eqnarray}}
\newcommand{\eea}{\end{eqnarray}}

\begin{document}
\title{Applicability of Self-Consistent Mean Field Theory}
\author{Lu Guo}
\affiliation {Department of Mathematical Science, Ibaraki University, Mito 310-8512, Ibaraki, Japan}
\author{Fumihiko Sakata}
\affiliation {Institute of Applied Beam Science, Graduate School of Science and Engineering, Ibaraki University, Mito 310-8512, Ibaraki, Japan} 
\author{En-guang Zhao}
\affiliation {Institute of Theoretical Physics, Chinese Academy of Sciences, Beijing 100080, China}
\affiliation {Department of Physics, Peking University, Beijing 100871, China}
\date{\today}

\begin{abstract}
Within the constrained Hartree-Fock (CHF) theory, an analytic condition is derived to estimate whether a concept of the self-consistent mean field is realized or not in level repulsive region. 
The derived condition states that an iterative calculation of CHF equation does not converge when the quantum fluctuations coming from two-body residual interaction and quadrupole deformation become larger than a single-particle energy difference between two avoided crossing orbits. 
By means of the numerical calculation, it is shown that the analytic condition works well for a realistic case.

\end{abstract}
\pacs{21.60.Jz, 21.10.Pc, 21.30.Fe}

\maketitle

\section{\label{level1} Introduction}

The Hartree-Fock (HF) mean field theory has been applied successfully to various quantum many-fermion systems.
By violating various conservation laws satisfied by the Hamiltonian, one obtains an appropriate mean field that could incorporate as many various correlations of the system as possible. 
Its many important theoretical concepts$-$such as a stability of the mean field and an appearance of massless collective motion in association with restoration dynamics of broken symmetries$-$constitute an outstanding virtue of the self-consistent mean field theory.
Here it should be noticed that the mean field is characterized by various averaged quantities, which express an amount of symmetry breaking like a shape deformation in the coordinate space or a gauge deformation in the quasi-spin (pairing) space.
These averaged quantities as well as the concept of mean field are expected to have physical reality only when the many-body system under consideration is near stationary, at least locally homogeneous, and well isolated from the other local minimum.
To the best of our knowledge, there has been no discussion of when a concept of the self-consistent mean field is realized or how and why it breaks down.
In this paper, we will discuss to what extent the self-consistent one-body potential exists when two orbits around the Fermi surface interact with each other.
 
Theoretically, an avoided crossing occurs rather frequently in various fields of physics like the molecular, atomic, biological systems, as well as the quantum dot and atomic nuclear systems \cite{JMA03,RGF03,FI01,CD01}.
In a development of the nuclear structure physics, there have been many discussions on an applicability of the cranked mean field theory near an avoided level crossing region \cite{Hama76,Stru77,Naz93,Dob00}, where large angular momentum fluctuations and spurious interactions between two crossing orbits have been explored.
An argument for removing a certain spurious interaction and for introducing a set of diabatic single particle (s.p.) states in the cranked mean field \cite{Naz93,Dob00,RB89,TB,RB,TB89,AA021,AA022} seems to be reasonable, because the angular momentum is a constant of motion and the interaction between different rotational bands should act at a given angular momentum rather than at a given rotational frequency. 
However, these treatments seem to be still in a phenomenological stage.

The above argument may not be simply extended to deformation constrained mean field theory.
In this case, an interaction between two potential energy surfaces (PESs) with different quadrupole deformation may not necessarily be regarded as spurious, because the Hamiltonian and quadrupole operator do not commute with each other.
Interestingly, it has been suggested that the pairing interaction between two orbits located below and above the Fermi surface and having spherical and deformed shape, might be spurious in the level crossing region \cite{RB89}.  By eliminating the spurious interactions and comparing with experiments, very interesting conclusions have been deduced such as "a flat PES does not automatically lead to large fluctuation of the shape" \cite{RB89}.  However, there still remains a decisive question of how the above statement is justified from the underlying dynamics. 

To understand how the mean field changes by itself when it acquires additional deformation, and to study more deeply what actually happens in the self-consistent mean field near the level crossing region from the underlying microscopic dynamics, a rather heavy numerical method called the configuration dictated CHF method \cite{CD1} has been developed.
Applying this method to realistic cases, one may get many PESs which are approximately characterized by the s.p. configurations relative to the lowest PES \cite{Guo3}. 
As pointed out in our previous papers \cite{Guo3, Guo2}, the CHF iterative calculation sometimes meets a difficulty of poor convergence or even nonconvergence near the level repulsive region, no matter how much efforts one makes to get convergence.

In this paper, we will explore how the competition between the mean field and the quantum fluctuations coming from two-body residual interaction and quadrupole deformation plays a decisive role in breaking the concept of a self-consistent mean field near the level repulsive region.

\section{\label{level2} theoretical method}
The nuclear Hamiltonian is given by
\beq
\hat H=\sum_{\alpha\beta}t_{\alpha \beta}c_{\alpha}^\dagger c_{\beta}+\frac{1}{4}\sum_{\alpha\beta\gamma\delta}\bar v_{\alpha\beta\gamma\delta}c_{\alpha}^\dagger c_{\beta}^\dagger c_{\delta}c_{\gamma}.
\eeq
Here and hereafter, $\alpha, \beta,\cdots$ are used to denote numerical basis. 
To define the self-consistent CHF state $|\Psi(q)\rangle$ with a given quadrupole moment $q$, we apply the deformation CHF equation with quadratic constraints given by
\bea
&&\delta \Bigl(\langle\Psi(q)|\hat{H}|\Psi(q)\rangle+\frac{1}{2}w\bigl(\langle\Psi(q)|\hat
Q_{20}|\Psi(q)\rangle-\mu\bigr)^2  \nonumber \\
&&+\frac{1}{2}\alpha_x \langle\Psi(q)|\hat x|\Psi(q)\rangle ^2\Bigr)=0,
\label{eq:chf1} 
\eea
with 
\beq
\langle\Psi(q)|\hat Q_{20}|\Psi(q)\rangle = q,\quad
\langle\Psi(q)|\hat x|\Psi(q)\rangle = 0,
\eeq
where $\hat x$ denotes a center of mass coordinate.
Our numerical method of solving the CHF equation has been discussed in Ref. \cite{Guo3}, where the Gogny D1S interaction \cite{Gog1,Gog2,Gog3,Gog4,Gog5,Gog6}, the Coulomb force and the center of mass motion up to the exchange terms are taken into account. 
In Eq.(\ref{eq:chf1}), $\mu$ denotes an input parameter that allows us to vary an expectation value $\langle\Psi(q)|\hat Q_{20}|\Psi(q)\rangle$. 
Meaning of parabola width $w$ was discussed in Ref. \cite{HFL}, and is chosen to be $1.0\times 10^{-3}{\textrm {MeV}}/{\textrm {fm}}^4$ in our calculation. 
The Lagrange multiplier $\lambda(q)$ is given by
\begin{equation} 
\lambda(q) = w\bigl(\mu-\langle\Psi(q)|\hat Q_{20}|\Psi(q)\rangle\bigr), 
\label{lamb}
\end{equation}
and an effective value of $\lambda(q)$ is allowed to change during the iterations.
The symmetries $\hat{P}e^{-i\pi\hat{J_z}}$ ($z$ simplex) and
$\hat{P}e^{-i\pi\hat{J_y}}\hat{\tau}$ ($\hat{S_y ^T}$) \cite{JD1,JD2} are imposed in our numerical calculation, where $\hat{P}$ is the parity operator, $e^{-i\pi\hat{J_i}}$ the rotation operator around {\it i} axis by an angle $\pi$, and $\hat\tau$ the time reversal operator. 
To keep the center of mass motion fixed, we impose a quadratic constraint in the $x$-axis direction with Lagrange multiplier 
\beq
\lambda^\prime =\alpha_{x}\langle\Psi(q)|\hat x|\Psi(q)\rangle,
\eeq
and take $\alpha_x=1.0\times10^{-4}{\textrm {MeV}}/{\textrm {fm}}^2$ in our numerical calculation. 

Having solved the CHF equation (\ref{eq:chf1}), one obtains a set of s.p. energies $\{\epsilon_{k}(q) \}$ as well as the s.p. states $\{ \varphi_{k}(q)\}$.
Hereafter, the particle states are denoted by $\mu$, $\nu$ and the hole states by $i, j$.
The letters $k, l$ are used when no distinction is needed.
To understand how the CHF state undergoes a structure change depending on the quadrupole deformation, it is desirable to obtain $|\Psi(q+\Delta q)\rangle$ in such a way that it can be regarded as a smooth function of $q$.  
For this aim, we apply the configuration dictated CHF method, which is briefly recapitulated below.
Let $|\Psi(q)\rangle$ be a known CHF state satisfying condition $\langle \Psi(q)| \hat Q_{20} |\Psi(q)\rangle =q$. 
To find a new CHF state $|\Psi(q+\Delta q)\rangle$ that is supposed to be continuously connected with $|\Psi(q)\rangle$, we exploit the following condition 
\begin{equation}
\lim_{\Delta q \rightarrow 0}\langle\varphi_i(q)|\varphi_j(q+\Delta q)\rangle=\delta_{i,j},
\label{cdm}
\end{equation}
where $\{\varphi_i(q)\}$ denotes a set of occupied wave functions constructing the single Slater determinant $|\Psi(q) \rangle$.
That is, a small increment $\Delta q$ is numerically adjusted by the maximum overlap criterion in Eq.~(\ref{cdm}) under a given accuracy, so as to maintain a characteristic property of the CHF state.
In our calculation, $\Delta q$ is so determined as to fulfill the condition 
\begin{equation}
|\langle\varphi_i(q)|\varphi_i(q+\Delta q)\rangle|^2 > 0.9. 
\label{cdm9}
\end{equation}
In this way, the configuration specifying $|\Psi(q)\rangle$ is kept continuously as a function of $q$.
Since the CHF state at $q+\Delta q$ is dictated by the configuration of the preceding CHF state at $q$, this method is called the configuration dictated CHF method. 
It can also be generalized to get the excited HF states and the continuously-connected PESs.
Applications of the configuration dictated constrained Hartree-Fock-Bogoliubov (CHFB) method in the level crossing region as well as in the shape coexistence phenomena, with and without the pairing effect have been reported elsewhere \cite{Guo3}.

In our numerical calculation, a convergence condition is given in eV as
\begin{equation}
\sum_{k}\left|\epsilon^{(n)}_k(q)-\epsilon^{(n-1)}_k(q)\right|\leq 10 ,
\label{eq12}
\end{equation}
where $\epsilon_k^{(n)}$ denotes the s.p. energies in the $n$th iteration.

\section{\label{level3} fragility of mean field}

\subsection{\label{level31} Difficulty of nonconvergence near the level repulsive region in CHF theory}

In our calculation, the s.p. wave functions are expanded on three-dimensional harmonic oscillator basis up to the principal quantum number $N_0=8$. 
The ground state of $^{66}{\textrm {Se}}$ is obtained after having optimized triaxial deformation parameters of the Hermite polynomials.
The optimized range parameters thus obtained include some effects of higher major shells. Note that, no optimization has been done in the most HF and HFB calculations \cite{ref1,ref2,ref3,ref4}, although a larger configuration space is adopted.
It is well known that the optimum parameters change depending on the number of major shell, and become rather stable as the number increases \cite{HFL}. In order to examine a reliability of the optimized configuration space, the ground state properties of $^{66}{\textrm {Se}}$ both in HF and HFB calculations are listed in Table~\ref{tab1}, together with the experimental binding energy \cite{Audi}. The binding energies in both calculations reproduce the experimental data well, though the total binding energy in the HFB is about 1.1 MeV lower than that in the HF calculation. 
A comparison between the HF and HFB calculations in Tab.~\ref{tab1} indicates that the nuclear deformation becomes small (favors the spherical shape) when one includes the pairing correlation into the mean field.
In the following discussion, the number of major shell used in our calculation may not have decisive importance.

\begin{table}
\caption{\label{tab1} The optimized ground state properties of $^{66}{\textrm {Se}}$.
The binding energy (BE), quadrupole deformation parameter ($\beta_2$) and triaxial deformation ($\gamma$) are listed.
The experimental BE is from Ref.~\cite{Audi}.}
\begin{ruledtabular}
\begin{tabular}{cccc}
        & HF  & HFB & Exp.  \\
\hline
BE [MeV]         & 544.502  &  545.623  &  547.827  \\
$\beta_2$        & 0.241    &  0.234    &            \\
$\gamma$ (degree)& 47.256   &  59.541   &
\end{tabular} 
\end{ruledtabular}
\end{table}

\begin{figure}
\epsfxsize=8.0cm
\centerline{\epsffile{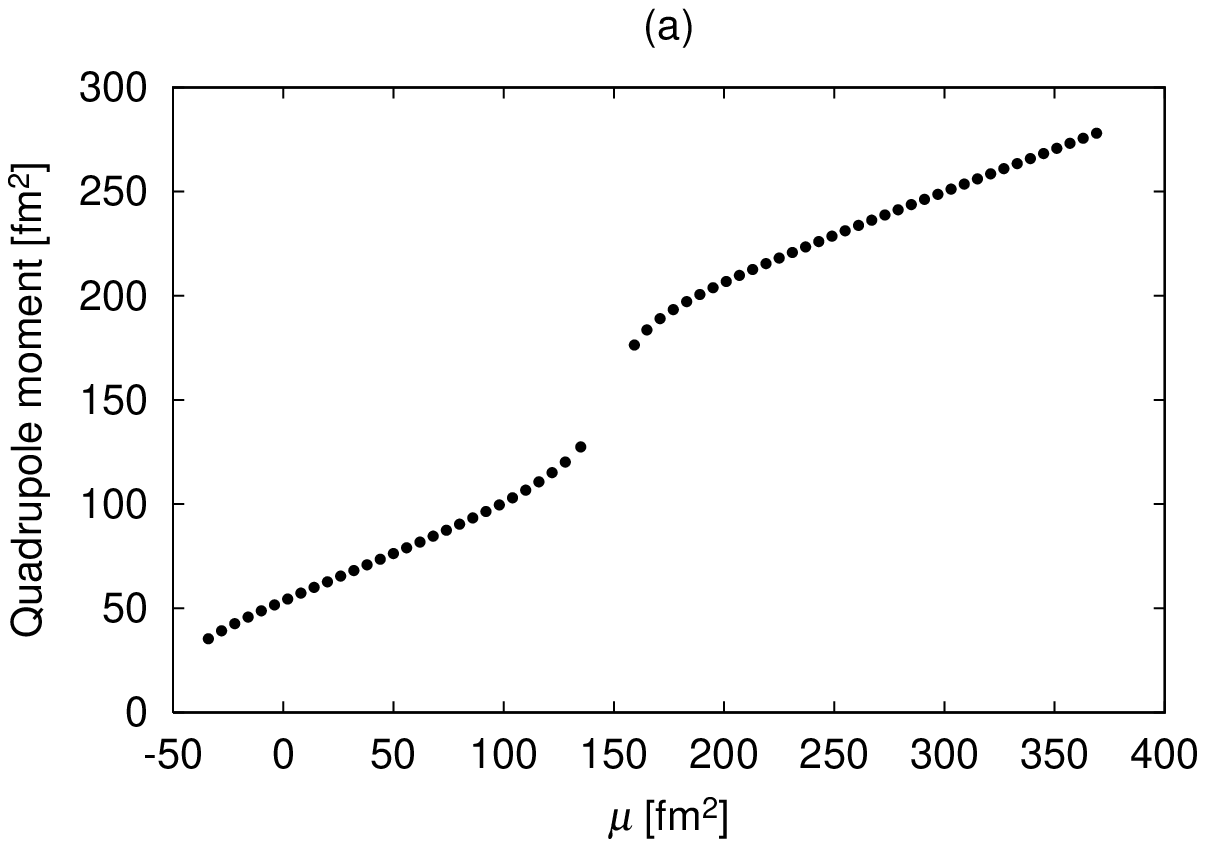}}
\centerline{\epsffile{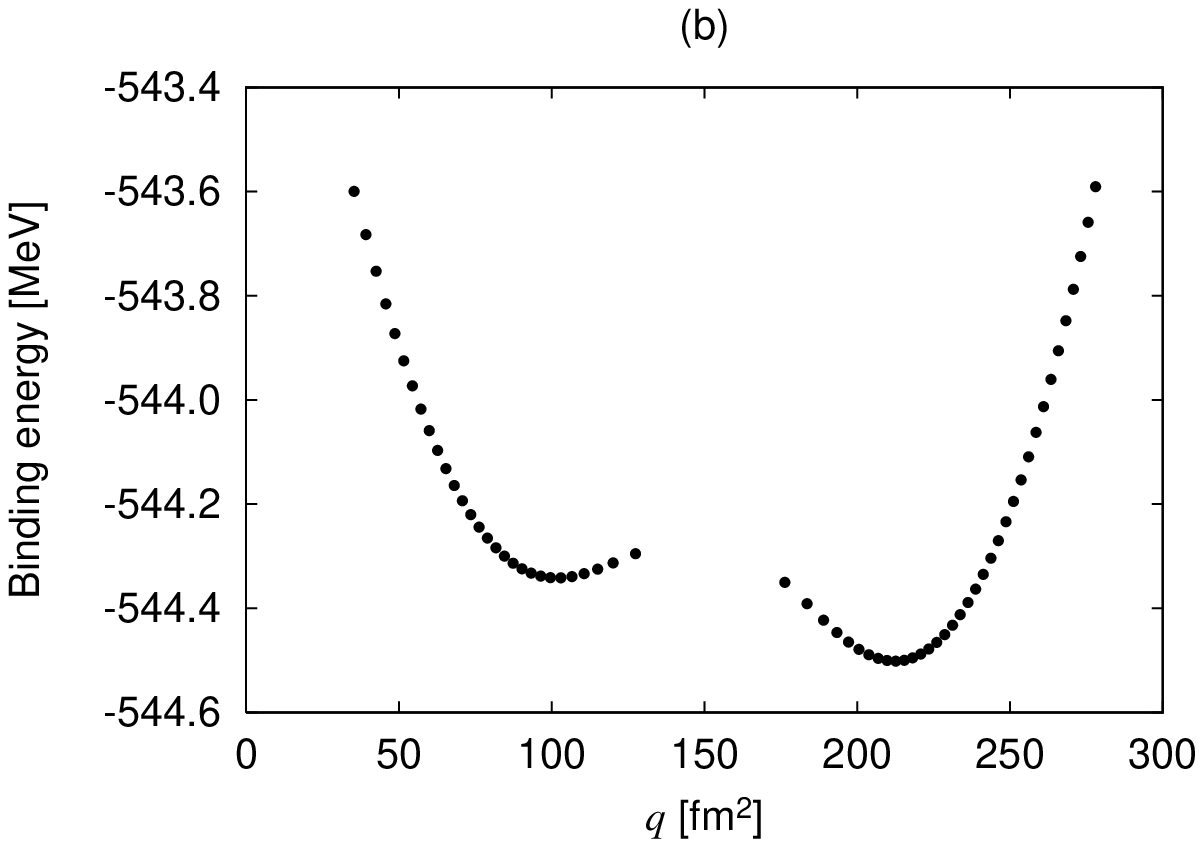}}
\caption{\label{Bind} CHF calculation for $^{66}{\textrm {Se}}$: (a) the calculated quadrupole moment as a function of input quadrupole moment parameter $\mu$; (b) binding energy as a function of quadrupole moment $q$. }
\end{figure}

Starting from the ground state, the quadratic CHF calculation with the configuration dictated method is carried out in such small steps as the solutions are considered to be a continuous function of the deformation. 
Such a point-by-point heavy calculation is needed for discussing the dynamical structure change of nuclear system. 
In numerically obtaining the PES, the same range parameters as those in the ground state are used to trace an evolution of the ground state configuration as a function of deformation, which makes the s.p. level crossing dynamics transparent. 
Figure~\ref{Bind} shows (a) the quadrupole moment as a function of $\mu$, and (b) the lowest PES for $^{66}{\textrm {Se}}$. One may observe that both the quadrupole deformation and PES change smoothly, except for a missing region around $q=150$ ${\textrm {fm}^2}$.
When the constrained quadrupole moment is decreased by a small amount $\Delta q$ from a critical point at $q_0=177.365$ ${\textrm {fm}^2}$, the CHF iteration meets a difficulty of nonconvergence no matter how much efforts one makes to get convergence.
After the missing region, the continuously-connected PES passing through an excited local minimum ($q \approx 100$ ${\textrm {fm}^2}$) is obtained. 

\begin{figure}
\epsfxsize=8.0cm
\centerline{\epsffile{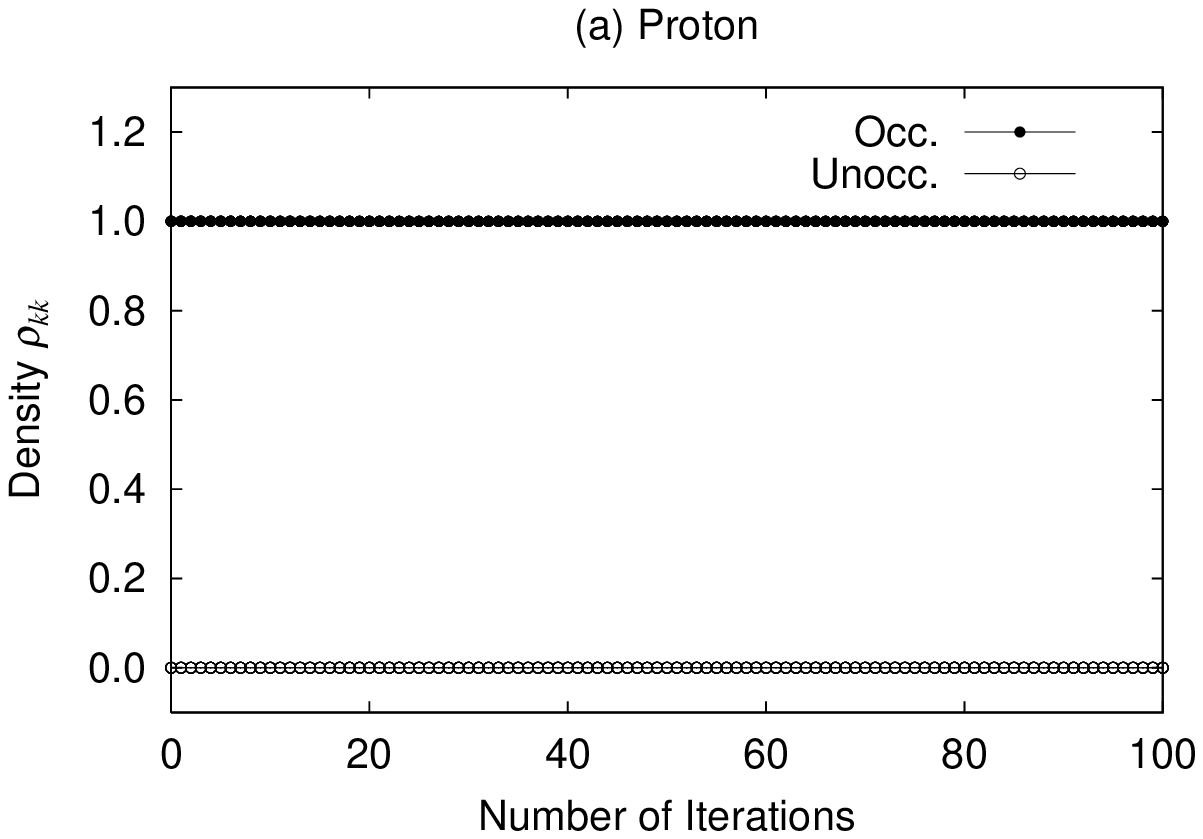}}
\centerline{\epsffile{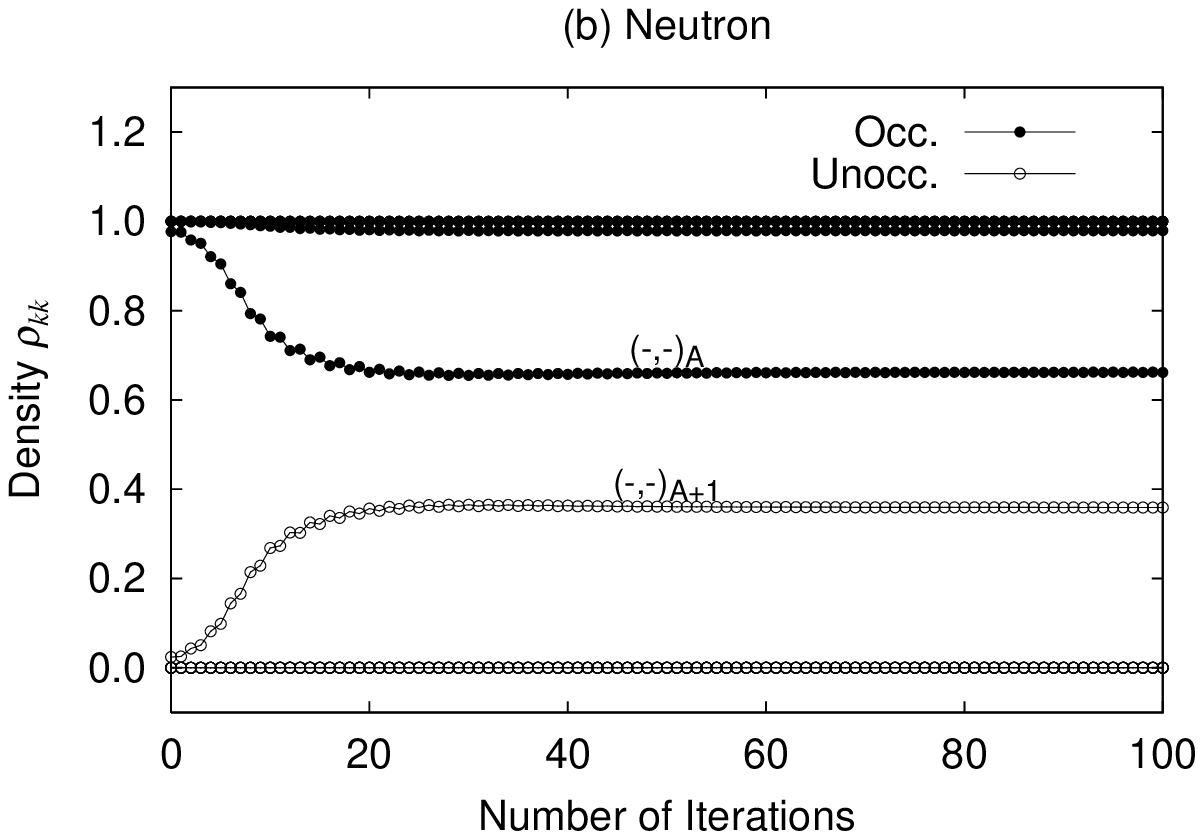}}
\caption{\label{Rhon} Diagonal components of (a) proton and (b) neutron density as a function of number of iteration for a nonconvergent case with $\mu=150$ ${\textrm {fm}^2}$. The single-particle basis where $h$ is diagonal is used. Occ. and Unocc. stand for occupied and unoccupied orbits, respectively. $(\pi,\alpha)$ denotes the parity and signature, and its subscripts $A$ and $A+1$ represent the orbits responsible for the nonconvergent difficulty.}
\end{figure}

The nonlinear CHF equation is solved in an iterative way until the CHF Hamiltonian $h(q)$ and density $\rho(q)$ are diagonalized simultaneously such that $[h(q), \rho(q)]=0$. 
 In the numerical basis, the s.p. wave function $\varphi_k^{(n)}(q)$ at the $n$th iteration is expressed as $\varphi_{\alpha k}^{(n)}(q)$, and $\rho^{(n)}(q)$ and $h^{(n)}(q)$ are given as 
\bea
\rho_{\alpha \beta}^{(n)}(q)&\equiv& \sum_i \varphi_{\alpha i}^{(n-1)}(q){\varphi_{\beta i}^{(n-1)}}^\ast(q), \nonumber \\
h_{\alpha \beta}^{(n)}(q)&\equiv& t_{\alpha \beta}+\Gamma_{\alpha\beta}^{(n)}(q) -\lambda^{(n)}(q)Q_{\alpha\beta}, \nonumber \\
\Gamma_{\alpha\beta}^{(n)}(q) &\equiv& \sum_{\gamma \delta}{\bar v}_{\alpha\gamma\beta\delta}\rho_{\delta \gamma}^{(n)}(q).\eea
Here $\Gamma(q)$ denotes the HF potential, and
\beq
\lambda^{(n)}(q)\equiv w\bigl(\mu-\langle\Psi^{(n-1)}(q)|\hat Q_{20}|\Psi^{(n-1)}(q)\rangle\bigr),
\eeq
where $|\Psi^{(n-1)}(q)\rangle$ is the Slater determinant constructed by $\{\varphi_i^{(n-1)}(q)\}$.

\begin{figure}
\epsfxsize=8.0cm
\centerline{\epsffile{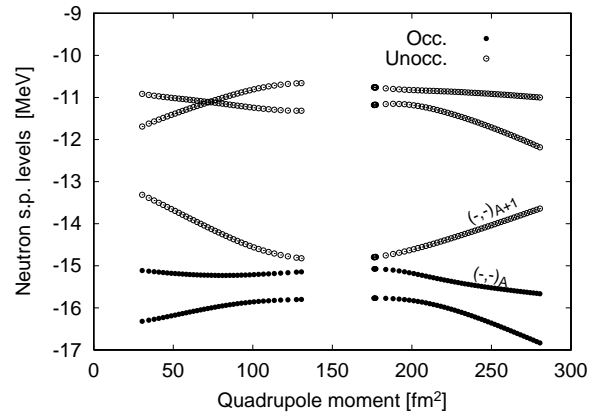}}
\caption{\label{Spln} Neutron single-particle energies near the Fermi surface. Occ. and Unocc. stand for occupied and unoccupied orbits, respectively. The symbols are the same as those in Fig.~\ref{Rhon}.}
\end{figure}

On a way to convergence, expectation values of one-body density $\rho^{(n)}(q)$ in a representation of using $\{ \varphi_{k}^{(n)}(q)\}$, i.e., 
\beq
\rho_{kk}^{(n)}(q)\equiv \sum_{\alpha\beta}{\varphi_{\alpha k}^{(n)}}^\ast(q)\rho_{\alpha \beta}^{(n)}(q)\varphi_{\beta k}^{(n)}(q),
\eeq
with respect to the hole and particle states are supposed to gradually reach 1 and 0, respectively.
To explain what prevents the CHF iterative calculation from convergence, Fig.~\ref{Rhon} depicts the diagonal components of proton and neutron densities as functions of number of iterations, for a case of nonconvergence with $\mu=150$ ${\textrm {fm}^2}$ (see Fig.~\ref{Bind}(a)). In the case of nonconvergence, the quadrupole deformation parameter $\mu$ is used in place of $q$.
One may observe that the expectation values of proton density for the unoccupied orbits converge to 0, while those for occupied orbits to 1. 
For the case of neutron, there appears a similar situation for the most single-hole and single-particle states, except for two specific orbits labeled as $(-,-)_A$ and $(-,-)_{A+1}$. 
Here, $A$ denotes the number of neutron-occupied orbits, and the s.p. states are specified by the parity and signature quantum numbers $(\pi,\alpha)$, since the asymptotic Nilsson quantum numbers are not good quantum numbers when the reflection symmetry is lost. 
For convenience, a subscript $A$ and $A+1$ for $(\pi,\alpha)$ is introduced to identify the occupied and unoccupied orbits responsible for the difficulty of nonconvergence.  
As seen from Fig.~\ref{Spln}, these two specific neutron orbits with opposite quadrupole moments, i.e., one is of deformation driving and the other of anti-driving, lying just below and above the Fermi surface are interacting. 
One may expect that the two specific orbits $(-,-)_A$ and $(-,-)_{A+1}$ play a dominant role in preventing the CHF calculation from convergence.

\begin{figure}
\epsfxsize=8.0cm
\centerline{\epsffile{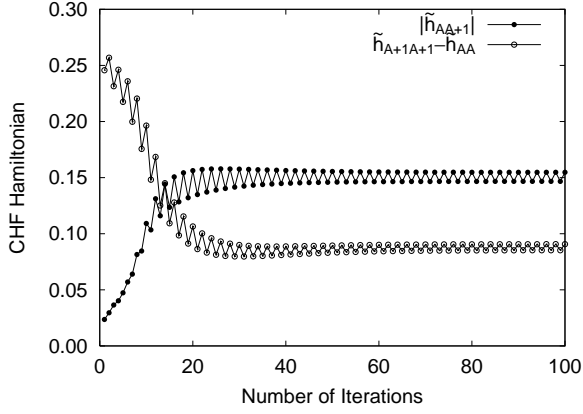}}
\caption{\label{hfhr} The absolute value of off-diagonal component and the difference of diagonal components of the CHF Hamiltonian between two specific orbits as a function of number of iteration for the nonconvergent case with $\mu=$150.0 ${\textrm {fm}}^2$. A representation of using s.p. basis where density $\rho^{(n)}$ is diagonal is used.}
\end{figure}

To exhibit crucial effects of two neutron orbits on the nonconvergence property in the CHF calculation, an absolute value of off-diagonal CHF Hamiltonian $| {\tilde h}_{A,A+1}^{(n)}|$ and a difference of diagonal components ${\tilde h}_{A+1,A+1}^{(n)}-{\tilde h}_{A,A}^{(n)}$ for the case with $\mu=$150.0 ${\textrm {fm}}^2$ are shown in Fig.~\ref{hfhr} as a function of number of iterations.
Here, the CHF Hamiltonian is expressed in a representation where density matrix $\rho^{(n)}$ is diagonal, i.e.,
\beq
{\tilde h}_{kl}^{(n)}(q) \equiv \sum_{\alpha\beta}{\varphi_{\alpha k}^{(n-1)}}^\ast(q)h_{\alpha\beta}^{(n)}(q)\varphi_{\beta l}^{(n-1)}(q).
\label{eq21}
\eeq
In the CHF theory, it is usual to employ the above representation where the $n$-th quantities are entirely expressed in terms of the $(n-1)$th s.p. wave function $\{\varphi_{\alpha k}^{(n-1)}\}$.

In a case of convergence, the off-diagonal component of the CHF Hamiltonian would become smaller and finally reach 0, as the number of iteration increased.
In the present nonconvergent case, there appears a staggering property in both the diagonal and off-diagonal components. 
During the iteration, $|{\tilde h}_{A,A+1}^{(n)}|$ increases first and then starts to oscillate around some central value. 
In this case, the off-diagonal component always remains and never reaches 0, rather than being included into the mean field. Figure~\ref{hfhr} indicates that the two interacting orbits make it difficult to apply the CHF mean field theory.

\subsection{Microscopic dynamics of nonconvergence}
This subsection discusses how the nonconvergent difficulty and the staggering property appear as a result of the microscopic dynamics.
Although the staggering phenomena in Fig.~\ref{hfhr} has been known as ``ping-pong''\cite{Dob00}, it should be noted that the following analytic understanding and underlying physics are given for the first time.
Namely, it is explored whether or not one gets a convergent CHF state at $q=q_0-\Delta q$ (a corresponding deformation parameter $\mu=\mu_0-\Delta\mu$), when there exists a convergent CHF state $|\Psi(q_0)\rangle$ at $q=q_0$ with corresponding set of self-consistent s.p. states $\{\epsilon_k(q_0), \varphi_k(q_0)\}$.
Since the CHF state $|\Psi(q_0)\rangle$ is used as an initial trial wave function to proceed the iterative calculation at $q=q_0-\Delta q$, 
the following relations hold:
\bea
&&h_{\alpha\beta}^{(0)}(q_0-\Delta q)=h_{\alpha\beta}(q_0), \nonumber \\
&&\varphi_{\alpha k}^{(0)}(q_0-\Delta q)=\varphi_{\alpha k}(q_0). 
\label{eq22}
\eea

To study dynamical change of the s.p. wave functions during the iterations, it turns out to be preferable to use a fixed representation, i.e., the $q_0$-{\it representation} of using the CHF s.p. states $\{\varphi_k(q_0)\}$, rather than the usual representation of using $n$-dependent s.p. wave functions $\{\varphi_k^{(n-1)}(q_0-\Delta q)\}$ as in Eq.~(\ref{eq21}).
The $n$th Hamiltonian in the $q_0$-representation is expressed as
\bea
&&h_{k l}^{(n)}(q_0-\Delta q) \equiv \sum_{\alpha\beta}\varphi_{\alpha k}^*(q_0)h_{\alpha\beta}^{(n)}(q_0-\Delta q)\varphi_{\beta l}(q_0) \nonumber \\
&&=t_{kl}+\Gamma_{kl}^{(n)}(q_0-\Delta q)-\lambda^{(n)}(q_0-\Delta q)Q_{kl} \nonumber \\
&& \quad\mbox{for}\quad n\neq 0, \nonumber \\
&&h_{k l}^{(0)}(q_0-\Delta q) = h_{kl}(q_0) \quad \mbox{for}\quad n= 0,
\label{eq23}
\eea
where the matrix elements of kinetic energy, deformation operator and the HF potential in the $q_0$-representation are defined as
\bea
t_{kl} &\equiv& \sum_{\alpha\beta}\varphi^{*}_{\alpha k}(q_0)t_{\alpha\beta}\varphi_{\beta l}(q_0),\nonumber \\
Q_{kl} &\equiv& \sum_{\alpha\beta}\varphi^{*}_{\alpha k}(q_0)Q_{\alpha\beta}\varphi_{\beta l}(q_0),\nonumber \\
\Gamma_{kl}^{(n)}(q_0-\Delta q) &\equiv&  \sum_{\alpha\beta}\varphi^{*}_{\alpha k}(q_0)\Gamma_{\alpha\beta}^{(n)}(q_0-\Delta q)\varphi_{\beta l}(q_0),\nonumber \\
\eea
and
\bea
&&\lambda^{(n)}(q_0-\Delta q) = w(\mu_0-\Delta\mu \nonumber \\ 
&& -\langle\Psi^{(n-1)}(q_0-\Delta q)|\hat Q_{20}|\Psi^{(n-1)}(q_0-\Delta q)\rangle), \nonumber \\ &&\quad \mbox{for}\quad n\neq 0, \nonumber \\
&&\lambda^{(0)}(q_0-\Delta q) = \lambda(q_0)= w(\mu_0-\langle\Psi(q_0)|\hat Q_{20}|\Psi(q_0)\rangle), \nonumber \\ &&\quad \mbox{for}\quad n= 0.
\label{eq25}
\eea

Making the following analytic understanding of our numerical results transparent and simple, we exploit such an approximate expression, i.e.,
\beq
\sum_{i=1}^{A-1}\varphi_{\alpha i}(q_0)\varphi_{\beta i}^{\ast}(q_0)\approx\sum_{i=1}^{A-1}\varphi_{\alpha i}^{(n)}(q_0-\Delta q){\varphi_{\beta i}^{(n)}}^\ast(q_0-\Delta q),
\label{assmp}
\eeq 
which states that a contribution to the mean field from the lowest $(A-1)$ number of hole-states is independent of small deformation change $\Delta\mu$ as well as number of iteration $n$. 
It turns out that Eq.~(\ref{assmp}) is well justified in our numerical calculation discussed in the previous subsection. 
Employing the above simplification, one may explore the nonconvergent dynamics governing the CHF iterative calculation in terms of a $2\times 2$ truncated CHF Hamiltonian expressed as 
\beq 
\left( \begin{array}{cc} h_{A,A}^{(n)} (q_0-\Delta q) & h_{A,A+1}^{(n)} (q_0-\Delta q)  \\ h_{A+1,A}^{(n)} (q_0-\Delta q) & h_{A+1,A+1}^{(n)} (q_0-\Delta q) \end{array} \right).
\label{2dimh}
\eeq
Although a set of eigen states $\{\varphi_k^{(n)}(q_0-\Delta q)\}$ is numerically obtained by diagonalizing the full CHF Hamiltonian $h^{(n)}(q_0-\Delta q)$, the characteristic feature of two interacting orbits is expected to be understood in terms of the truncated $2\times 2$ Hamiltonian in Eq.~(\ref{2dimh}) and $2\times 2$ unitary matrix $U^{(n)}$ given by
\bea
\left( \begin{array}{c} \varphi_A ^{(n)}(q_0-\Delta q) \\  \varphi_{A+1} ^{(n)}(q_0-\Delta q) \end{array} \right) 
&=&U^{(n)} \left( \begin{array}{c} \varphi_A (q_0)\\  \varphi_{A+1}(q_0) \end{array} \right),\nonumber \\ 
U^{(n)} &=& \left( \begin{array}{cc} a^{(n)} & b^{(n)}  \\ d^{(n)} & c^{(n)}  \end{array} \right).
\label{trans1}
\eea
In accordance with using the $q_0$-representation, the above unitary transformation allows us to study the iterative process in terms of a mixing between two fixed states $\varphi_A(q_0)$ and $\varphi_{A+1}(q_0)$ irrespective of $n$. 
To explore a decisive role of relative phase between two orbits on the nonconvergent difficulty of the CHF iterative calculation, i.e., on the properties of resultant s.p. wave functions obtained after having diagonalizing $h_{k l}^{(n)} (q_0-\Delta q)$, we use four inter-dependent parameters in $U^{(n)}$ rather than a single independent parameter. Mixing parameters $b^{(n)}$ and $d^{(n)}$ (${b^{(n)}}^2={d^{(n)}}^2$) are a measure of the degree of mixing between the two specific orbits. 
In each diagonalization, the subscripts $A$ and $A+1$ are used to assign the s.p. states in an energy-increasing order satisfying $\epsilon_A^{(n)}(q_0-\Delta q)< \epsilon_{A+1}^{(n)}(q_0-\Delta q)$.

With the aid of Eq.~(\ref{trans1}) and the approximate expression in Eq.~(\ref{assmp}), it is easily shown that the Lagrange multiplier $\lambda^{(n)}(q_0-\Delta q)$ in Eq.~(\ref{eq25}) fulfills the following simple recurrence relation
\begin{widetext}
\beq
\lambda^{(n+1)}(q_0-\Delta q)=\lambda^{(n)}(q_0-\Delta q)+\Delta\lambda^{(n)}(q_0-\Delta q)\quad \mbox{for}\quad n\neq 0,
\eeq
where
\beq
\Delta\lambda^{(n)}(q_0-\Delta q)= -2w{a^{(n)}}{b^{(n)}} Q_{A,A+1}-w{b^{(n)}}^2\{ Q_{A+1,A+1}-Q_{A,A}\} \quad \mbox{for}\quad n\neq 0 ,
\label{eqlag}
\eeq
with an initial relation given as
\bea
\lambda^{(1)}(q_0-\Delta q)&=&w(\mu_0-\Delta \mu-\langle\Psi^{(0)}(q_0-\Delta q)|\hat Q_{20}|\Psi^{(0)}(q_0-\Delta q)\rangle)\nonumber \\
&=&\lambda^{(0)}(q_0-\Delta q)-w \Delta \mu=\lambda(q_0)-w \Delta \mu \quad \mbox{for}\quad n= 0.
\label{eq31}
\eea
In the same way, one may derive the general expression of the matrix elements of $2\times 2$ truncated CHF Hamiltonian in Eq.~(\ref{2dimh}) at the $(n+1)$th iteration as 
\bea
h_{A,A+1}^{(n+1)}&=&h_{A,A+1}^{(1)}+{a^{(n)}}{b^{(n)}}(\bar v_{A+1AAA+1}+2wQ_{A,A+1}^2)+w{b^{(n)}}^2 Q_{A,A+1}(Q_{A+1,A+1}-Q_{A,A}),\nonumber \\
h_{A,A}^{(n+1)}&=&h_{A,A}^{(1)}+2w{a^{(n)}}{b^{(n)}}Q_{A,A}Q_{A,A+1} 
-{b^{(n)}}^2\bigl\{\bar v_{A+1AAA+1}-wQ_{A,A}(Q_{A+1,A+1}-Q_{A,A})\bigl\},\nonumber \\
h_{A+1,A+1}^{(n+1)}&=&h_{A+1,A+1}^{(1)}+2w{a^{(n)}}{b^{(n)}}Q_{A+1,A+1}Q_{A,A+1}
+{b^{(n)}}^2 \bigl\{\bar v_{A+1AAA+1}+wQ_{A+1,A+1}(Q_{A+1,A+1}-Q_{A,A})\bigl\},\nonumber \\ 
&& \quad \mbox{for}\quad n\neq 0,
\label{hfh3}
\eea  
\end{widetext}
where the anti symmetrized two-body interaction in the $q_0$ representation is defined as
\beq
\bar v_{k_1k_2k_3k_4} \equiv \sum_{\alpha\beta\gamma\delta}\varphi_{\alpha k_1}^{*}(q_0)\varphi_{\gamma k_2}^{*}(q_0) \bar v_{\alpha\gamma\beta\delta} \varphi_{\beta k_3}(q_0)\varphi_{\delta k_4}(q_0).
\eeq
Here and hereafter, we use a simple notation $h^{(n+1)}$ instead of $h^{(n+1)}(q_0-\Delta q)$ etc, because we are only discussing the CHF iterative process at $q=q_0-\Delta q$.
The matrix elements of truncated Hamiltonian $h_{kl}^{(1)}$ are given as
\bea
h_{A,A+1}^{(1)}&=&w\Delta\mu Q_{A,A+1},\nonumber\\
h_{A,A}^{(1)}&=&\epsilon_A(q_0)+w\Delta\mu Q_{A,A},\nonumber\\
h_{A+1,A+1}^{(1)}&=&\epsilon_{A+1}(q_0)+w\Delta\mu Q_{A+1,A+1},
\label{hfh1}
\eea
which are easily derived by using the initial condition in Eq.~(\ref{eq22}) and initial relations in Eqs.~(\ref{eq23}), (\ref{eq25}) and (\ref{eq31}) for the case with $n=0$. 
In deriving Eq.~(\ref{hfh1}), the relation
\beq
\left( \begin{array}{cc} h_{A,A}^{(0)} & h_{A,A+1}^{(0)}  \\ h_{A+1,A}^{(0)} &  h_{A+1,A+1}^{(0)} \end{array} \right)
=\left( \begin{array}{cc} \epsilon_A(q_0) & 0 \\ 0 &  \epsilon_{A+1}(q_0) \end{array} \right),
\eeq
which is satisfied by the preceding CHF solution at $q=q_0$, is also used.

With the aid of Eq.~(\ref{hfh3}), we get the following relations for two successive truncated CHF Hamiltonians $h^{(n)}$ and $h^{(n+1)}$:
\begin{widetext}
\bea
h_{A,A+1}^{(n+1)}
&=&h_{A,A+1}^{(n)}+\{{a^{(n)}}{b^{(n)}}-{a^{(n-1)}}{b^{(n-1)}}\}\{\bar v_{A+1AAA+1}+2wQ_{A,A+1}^2\}  \nonumber \\
&+&w\{{b^{(n)}}^2-{b^{(n-1)}}^2\} Q_{A,A+1} (Q_{A+1,A+1}-Q_{A,A}) ,\nonumber \\
h_{A,A}^{(n+1)}
&=&h_{A,A}^{(n)}+2w\{{a^{(n)}}{b^{(n)}}-{a^{(n-1)}}{b^{(n-1)}}\}Q_{A,A} Q_{A,A+1}\nonumber \\
&-&\{{b^{(n)}}^2-{b^{(n-1)}}^2\}\{\bar v_{A+1AAA+1}-w Q_{A,A}( Q_{A+1,A+1}- Q_{A,A})\},\nonumber \\
h_{A+1,A+1}^{(n+1)}
&=&h_{A+1,A+1}^{(n)}+2w\{{a^{(n)}}{b^{(n)}}-{a^{(n-1)}}{b^{(n-1)}}\}Q_{A+1,A+1} Q_{A,A+1}\nonumber \\
&+&\{{b^{(n)}}^2-{b^{(n-1)}}^2\}\{\bar v_{A+1AAA+1}+w Q_{A+1,A+1}( Q_{A+1,A+1}- Q_{A,A})\}.
\label{eq34}
\eea
In the same way, the off-diagonal component of the difference between $h^{(n)}$ and $h^{(n+2)}$ is given as
\bea
h_{A,A+1}^{(n+2)}
&=&h_{A,A+1}^{(n)}+\{{a^{(n+1)}}{b^{(n+1)}}-{a^{(n-1)}}{b^{(n-1)}}\}\{\bar v_{A+1AAA+1}+2wQ_{A,A+1}^2\}\nonumber \\
&+&w\{{b^{(n+1)}}^2-{b^{(n-1)}}^2\} Q_{A,A+1} (Q_{A+1,A+1}-Q_{A,A}).
\label{ite3}
\eea
\end{widetext}
Since the second term in the right-hand side of Eq.~(\ref{eq34}) contains a factor $a^{(n)}b^{(n)}$ whereas the third term has a factor $b^{(n)2}$, the former term is retained in the following discussions because the parameter $b^{(n)}$ is small.

\begin{figure}
\epsfxsize=8.0cm
\centerline{\epsffile{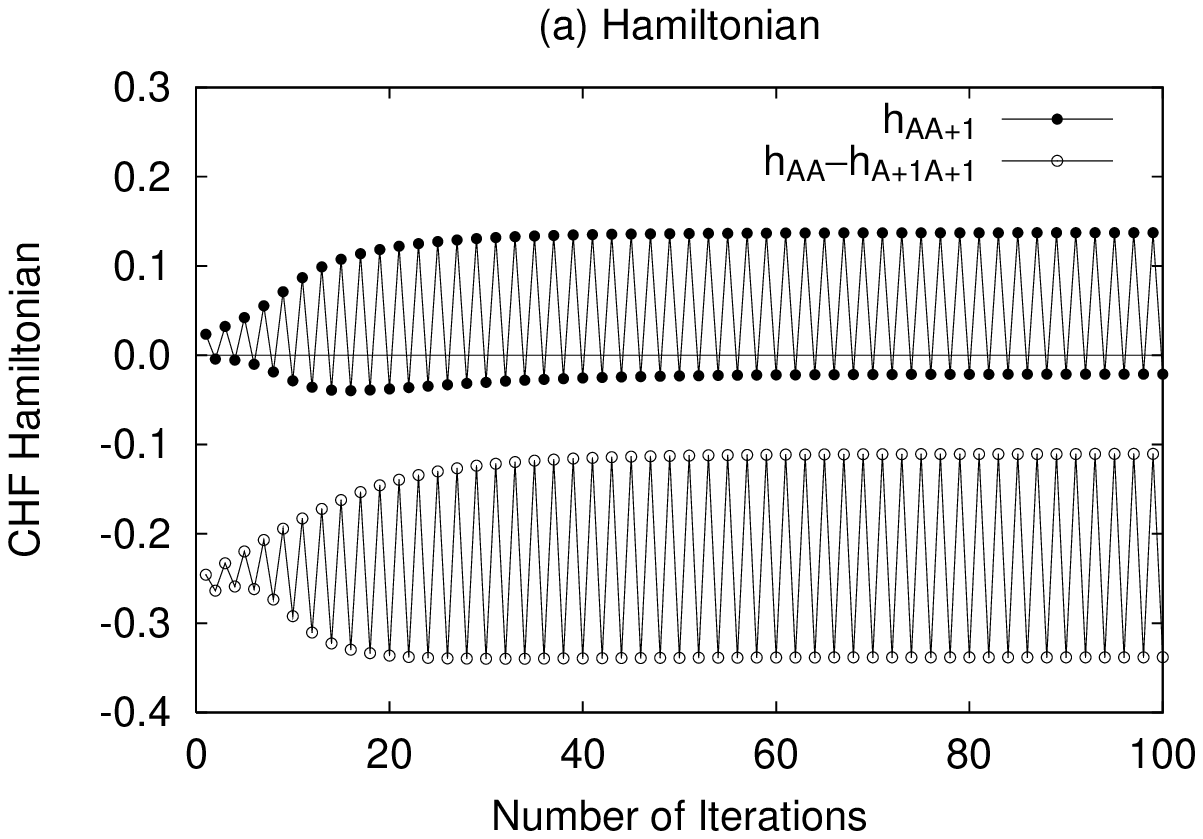}}
\centerline{\epsffile{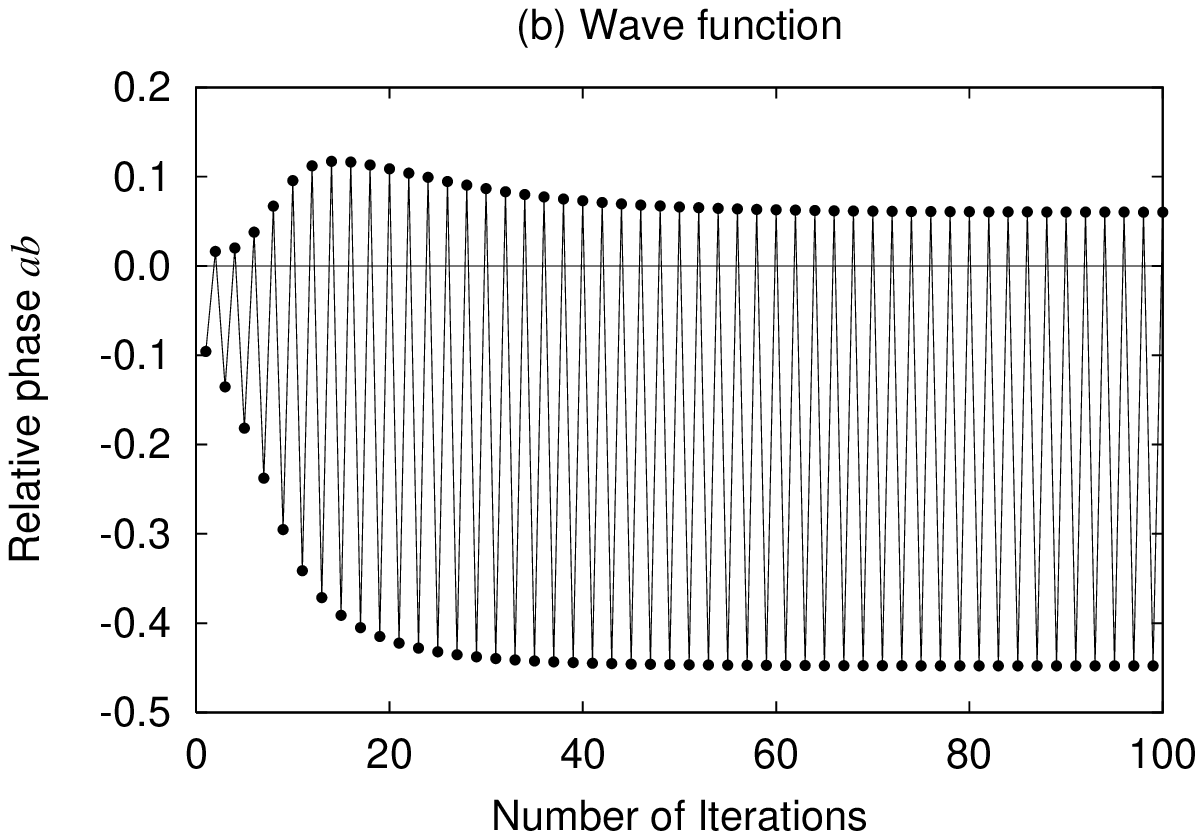}}
\caption{\label{Hfh} For the nonconvergent case with $\mu=150$ ${\textrm {fm}^2}$
(a) the off-diagonal Hamiltonian and difference of diagonal components as a function of number of iteration;
(b) the relative phase $a^{(n)}b^{(n)}$ of wave function for the orbit $\varphi_A^{(n)}$.}
\end{figure}

\begin{figure}
\epsfxsize=8.0cm
\centerline{\epsffile{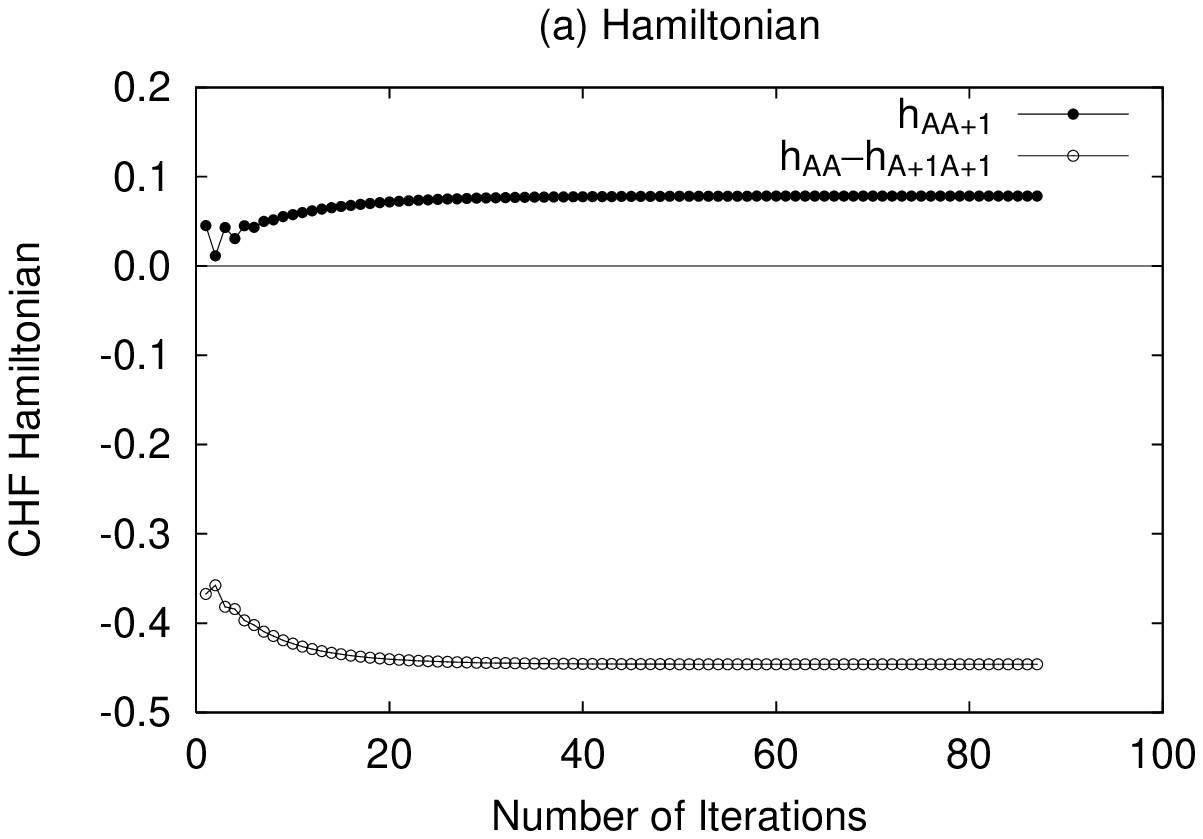}}
\centerline{\epsffile{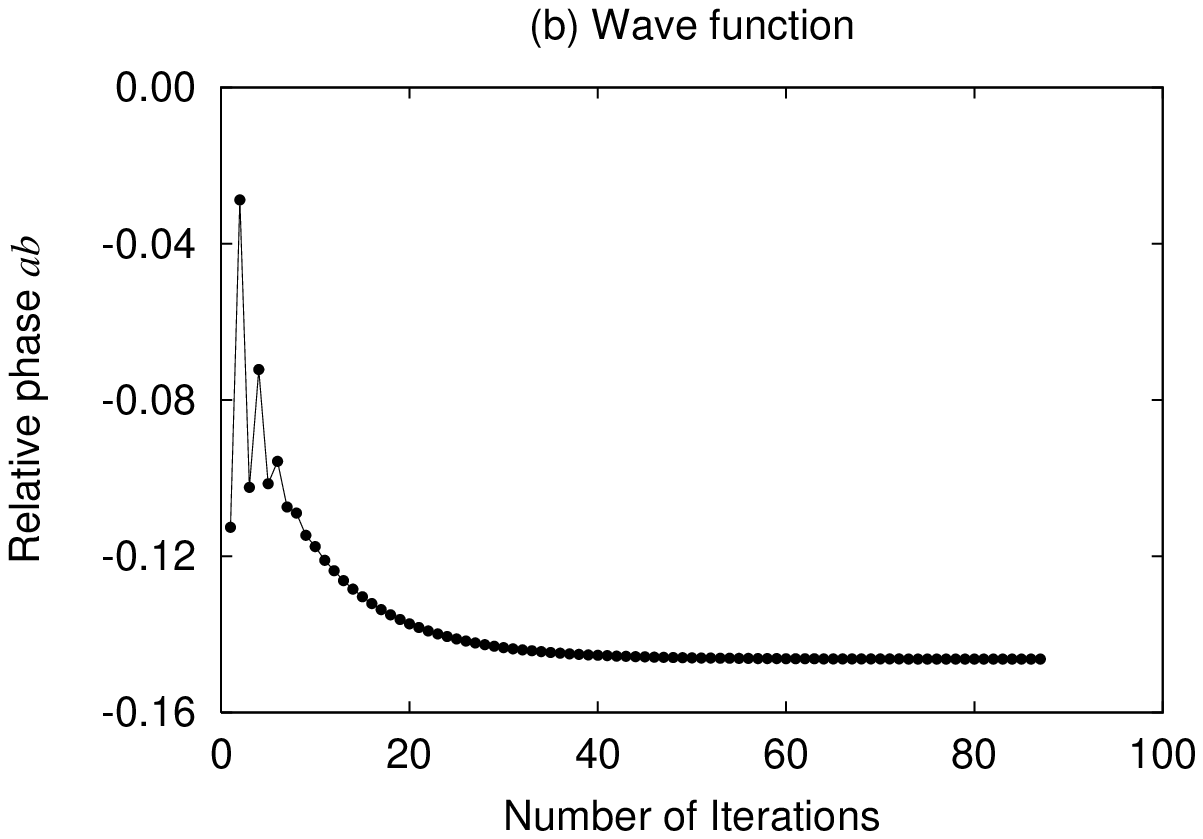}}
\caption{\label{Hfhc} For the convergent case with $\mu=180$ ${\textrm {fm}^2}$. Notation is the same as Fig.~\ref{Hfh}.}
\end{figure}

Numerical values of $ h^{(n)}_{A,A} - h^{(n)}_{A+1,A+1}$ and $ h^{(n)}_{A,A+1}$ are shown in Fig.~\ref{Hfh}(a) for the nonconvergent case with $\mu=150$ ${\textrm {fm}^2}$. 
From this figure, one may observe that both components exhibit a staggering property around some averaged values. 
Notice also that the off-diagonal component $h^{(n)}_{A,A+1}$ changes its sign from iteration to iteration, whereas $h_{AA}^{(n)}-h_{A+1 A+1}^{(n)}$ is always negative.
The latter is easily understood because a couple of states $\varphi_A^{(n)}(q-\Delta q)$ and $\varphi_{A+1}^{(n)}(q-\Delta q)$ is defined so as to satisfy a relation $\epsilon_A^{(n)}(q_0-\Delta q)< \epsilon_{A+1}^{(n)}(q_0-\Delta q)$.
According to the perturbation theory of the $2\times 2$ Hamiltonian,  the sign of $a^{(n)} b^{(n)}$ for characterizing the lower state $\varphi_A^{(n)}(q_0-\Delta q)$ is the same as that of $(h^{(n)}_{A,A+1})/(h^{(n)}_{A,A}-h^{(n)}_{A+1,A+1})$. 
In Fig.~\ref{Hfh}(b), the numerical value of $a^{(n)} b^{(n)}$ is shown as a function of number of iterations.
By comparing Fig.~\ref{Hfh}(a) with Fig.~\ref{Hfh}(b), our numerical results are well understood in terms of the above simplified analytic expression in the two dimensional truncated space and the perturbation theory for the $2\times 2$ Hamiltonian. 
Since the sign of $a^{(n)} b^{(n)}$ changes from iteration to iteration, it is clear that the higher state $\varphi_{A+1}^{(n)}(q_0-\Delta q)$ and the lower state $\varphi_A^{(n)}(q_0-\Delta q)$ inter-change their properties from iteration to iteration. 

With the aid of coefficient $a^{(n)}b^{(n)}-a^{(n-1)}b^{(n-1)}$ in the second term of the right-hand side of Eq.~{(\ref{eq34})} and from the above discussion on different signs between $a^{(n)}b^{(n)}$ and $a^{(n-1)}b^{(n-1)}$, one may understand why $h_{A,A+1}^{(n+1)}$ becomes large. 
Since there is a coefficient $a^{(n+1)}b^{(n+1)}-a^{(n-1)}b^{(n-1)}$ in the second term of the right-hand side of Eq.~{(\ref{ite3})}, and since the signs of $a^{(n+1)}b^{(n+1)}$ and $a^{(n-1)}b^{(n-1)}$ are the same, it is easily recognized  why the $ph$ component at $(n+2)$th iteration becomes small. 
These equations clearly state that the quantum fluctuations coming from two-body residual interaction and quadrupole deformation become small in one iteration and become large in the next, forming the staggering property.
In other words, the major part of the two-body interaction could not be approximated successfully by the averaged one-body potential when the sign of $a^{(n)}b^{(n)}$ changes from one iteration to the next. 
Physically one may understand the above situation as follows: two mean fields, one characterized by occupied $\varphi_A(q_0)$ and the other by occupied $\varphi_{A+1}(q_0)$, interact too strongly by the two-body residual interaction to be approximated by a single mean field.

In contrast with the nonconvergent case, the same quantities for convergent CHF calculation at $\mu=180$ ${\textrm {fm}^2}$ are shown in Fig.~\ref{Hfhc}. One may observe that the sign of off-diagonal Hamiltonian stays the same during the iteration, and the sign of $a^{(n)}b^{(n)}$ is always the same during the iteration until the convergence is achieved.
Since the sign of $a^{(n)}b^{(n)}$ does not change during the iterations in the case of convergence, the quantum fluctuations characterized by the coefficient ${a^{(n)}}{b^{(n)}}-{a^{(n-1)}}{b^{(n-1)}}$ in Eq.~{(\ref{eq34})} and by ${a^{(n+1)}}{b^{(n+1)}}-{a^{(n-1)}}{b^{(n-1)}}$ in Eq.~{(\ref{ite3})} become small as $n$ increases. That is, the quantum fluctuations are successfully incorporated into the mean field during the iterations. 
The above analytic and numerical results indicate clearly that the sign of $a^{(n)}b^{(n)}$ between two successive iterations is important in determining  whether the self-consistent CHF theory would be applicable.

\section{analytic conditions on applicability of mean field theory}

Since the analytic formulas within the two dimensional subspace discussed in the previous section nicely explain the characteristic feature of numerical iterative process of the CHF calculation near the level repulsive region, our next task is to determine what condition is capable of estimating an existence of the mean field.
In the previous section, it is clarified that the CHF theory breaks down when the off-diagonal Hamiltonian $h_{A,A+1}^{(n)}$ changes its sign from iteration to iteration, whereas it is successfully applied when the sign is kept the same during the iteration. 
Namely, it is shown that the ratio of two successive off-diagonal components $h_{A,A+1}^{(n+1)}/h_{A,A+1}^{(n)}$, [especially the first relation, $h_{A,A+1}^{(2)}/h_{A,A+1}^{(1)}$] plays a decisive role in the existence of the mean field.
That is, the self-consistent mean field exists when 
\beq
\left\{ \frac{h_{A,A+1}^{(2)}}{h_{A,A+1}^{(1)}}\right\} \ge 0.
\label{condition}
\eeq
From the discussion in the previous section, it is easily shown that the sign of $h_{A,A+1}^{(n+1)}/h_{A,A+1}^{(n)}$ is always positive when the condition in Eq.(\ref{condition}) is satisfied, while it changes its sign when the condition is not satisfied.

In this section, we focus on  discussing what is meant by the condition (\ref{condition}).
The ratio of two successive off-diagonal components satisfies the relation
\bea
\mbox{sign}\left\{ \frac{h_{A,A+1}^{(n+1)}}{h_{A,A+1}^{(n)}}\right\}&=&\mbox{sign}\left\{\frac{h_{A,A+1}^{(n+1)}/a^{(n)}b^{(n)}}{h_{A,A+1}^{(n)}/a^{(n)}b^{(n)}}\right\} \nonumber \\
&=&-\mbox{sign}\left\{ h_{A,A+1}^{(n+1)}/a^{(n)}b^{(n)} \right\},
\label{ratio}
\eea
because the signs of $h_{A,A+1}^{(n)}$ and $a^{(n)}b^{(n)}$ are always opposite, which is derived from our choice of states satisfying $\epsilon_A^{(n)}(q_0-\Delta q)< \epsilon_{A+1}^{(n)}(q_0-\Delta q)$ and from the discussion based on the perturbation theory and is justified by our numerical calculation.
With the aid of Eq.(\ref{hfh3}), the last quantity in (\ref{ratio}) is expressed as
\begin{widetext}
\beq
\frac{h_{A,A+1}^{(n+1)}}{a^{(n)}b^{(n)}}
=\frac{h_{A,A+1}^{(1)}+{a^{(n)}}{b^{(n)}}(\bar v_{A+1AAA+1}+2wQ_{A,A+1}^2)+w{b^{(n)}}^2 Q_{A,A+1}(Q_{A+1,A+1}-Q_{A,A})}{a^{(n)}b^{(n)}}.
\label{ratiox}
\eeq
\end{widetext}

Since an applicability of the mean field is supposed to be decided by a competition between the quantum fluctuation ($ph$ component of $h^{(1)}$) and the mean field part ($pp$ and $hh$ components of  $h^{(1)}$), let us find their relation by using the following relation
\beq
{U^{(1)}} h^{(1)} U^{(1)\dagger} 
= \left( \begin{array}{cc} \epsilon_A^{(1)} & 0  \\ 0 & \epsilon_{A+1}^{(1)}  \end{array} \right),
\eeq
which should approximately hold within the truncated space under the assumption in Eq.~(\ref{assmp}).
From an off-diagonal component of the above relation, we obtain
\beq
h_{A+1,A+1}^{(1)}-h_{A,A}^{(1)}=\Bigl\{\frac{c^{(1)}}{d^{(1)}}+\frac{b^{(1)}}{a^{(1)}}\Bigl\}h_{A,A+1}^{(1)}.
\label{rela1}
\eeq
Inserting Eq.~(\ref{hfh1}) for (\ref{rela1}) produces the relation
\begin{widetext}
\beq
h_{A,A+1}^{(1)} = \left\{\epsilon_{A+1}(q_0)-\epsilon_A(q_0)+w\Delta\mu(Q_{A+1,A+1}-Q_{A,A})\right\}/\left\{ \frac{c^{(1)}}{d^{(1)}}+\frac{b^{(1)}}{a^{(1)}}\right\}.
\label{rela2}
\eeq
With the aid of {(\ref{rela2})},  a quantity in {(\ref{ratiox})} for the case with $n=1$ is expressed as
\beq
\frac{h_{A,A+1}^{(2)}}{a^{(1)}b^{(1)}}
=\frac{1}{b^{(1)2}-a^{(1)2}} \left\{
\begin{array}{l}
 \epsilon_{A+1}(q_0)-\epsilon_{A}(q_0)-( \bar v_{A+1AAA+1}+2wQ^2_{A,A+1}) \\
 -w\Delta\mu (Q_{A,A}-Q_{A+1,A+1})-O(b) \\
\end{array}
\right\},
\eeq
where the higher-order term containing a small parameter $b^{(1)}$ is written as
\bea
O(b)=-2{b^{(1)}}^{2}\bigl(\bar v_{A+1AAA+1}+2wQ_{A,A+1}^2\bigl) 
+\frac{wb^{(1)}}{a^{(1)}}\bigl(1-2{b^{(1)}}^2\bigl)Q_{A,A+1}\bigl(Q_{A+1,A+1}-Q_{A,A}\bigl). \nonumber \\ 
\eea
\end{widetext}
Since the factor $({b^{(1)2}-a^{(1)2}})$ is always negative [because the relation $|a^{(1)}| \ge |b^{(1)}|$ holds in the unitary transformation in (\ref{trans1})), one finally gets a relation,
\bea
&&\mbox{sign}\left\{ \frac{h_{A,A+1}^{(2)}}{h_{A,A+1}^{(1)}}\right\}=
-\mbox{sign}\left\{ \frac{h_{A,A+1}^{(2)}}{a^{(1)}b^{(1)}}\right\}\nonumber \\
&&=\mbox{sign}\{
\epsilon_{A+1}(q_0)-\epsilon_{A}(q_0)-( \bar v_{A+1AAA+1}+2wQ_{A,A+1}^2)\nonumber \\
&& -w\Delta\mu (Q_{A,A}-Q_{A+1,A+1})-O(b) \}.
\eea
With the aid of this relation, our condition (\ref{condition}) is expressed as
\bea
\epsilon_{A+1}(q_0)-\epsilon_{A}(q_0) & \ge  & \bar v_{A+1AAA+1}+2w{Q^2_{A,A+1}} \nonumber \\
&+& w\Delta\mu (Q_{A,A}-Q_{A+1,A+1})+O(b), \nonumber \\
\label{con}
\eea
which guarantees an existence of the self-consistent mean field, whereas the opposite condition,
\bea
\epsilon_{A+1}(q_0)-\epsilon_{A}(q_0) & < &  \bar v_{A+1AAA+1}+2w{Q^2_{A,A+1}}\nonumber \\
&+& w\Delta\mu (Q_{A,A}-Q_{A+1,A+1})+O(b), \nonumber \\
\label{con1}
\eea
states a breakdown of the mean field. 
Note that the sign of the two-body interaction is important in Eqs.(\ref{con}) and (\ref{con1}), which is in mark contrast with the well-known stability condition of the mean field.
Furthermore, the present condition on a breakdown of the mean field has single-particle character, whereas the stability condition of the mean field has collective character.

\begin{figure}
\epsfxsize=8.0cm
\centerline{\epsffile{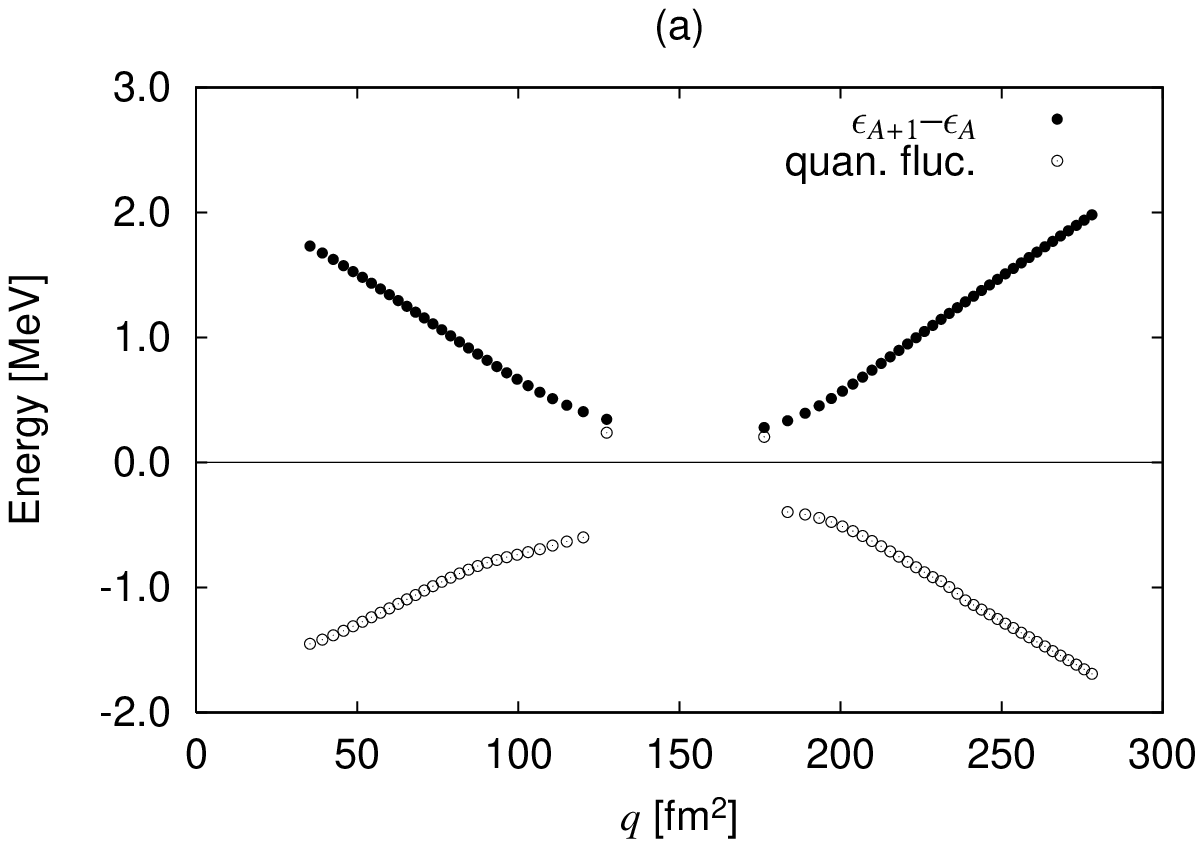}}
\centerline{\epsffile{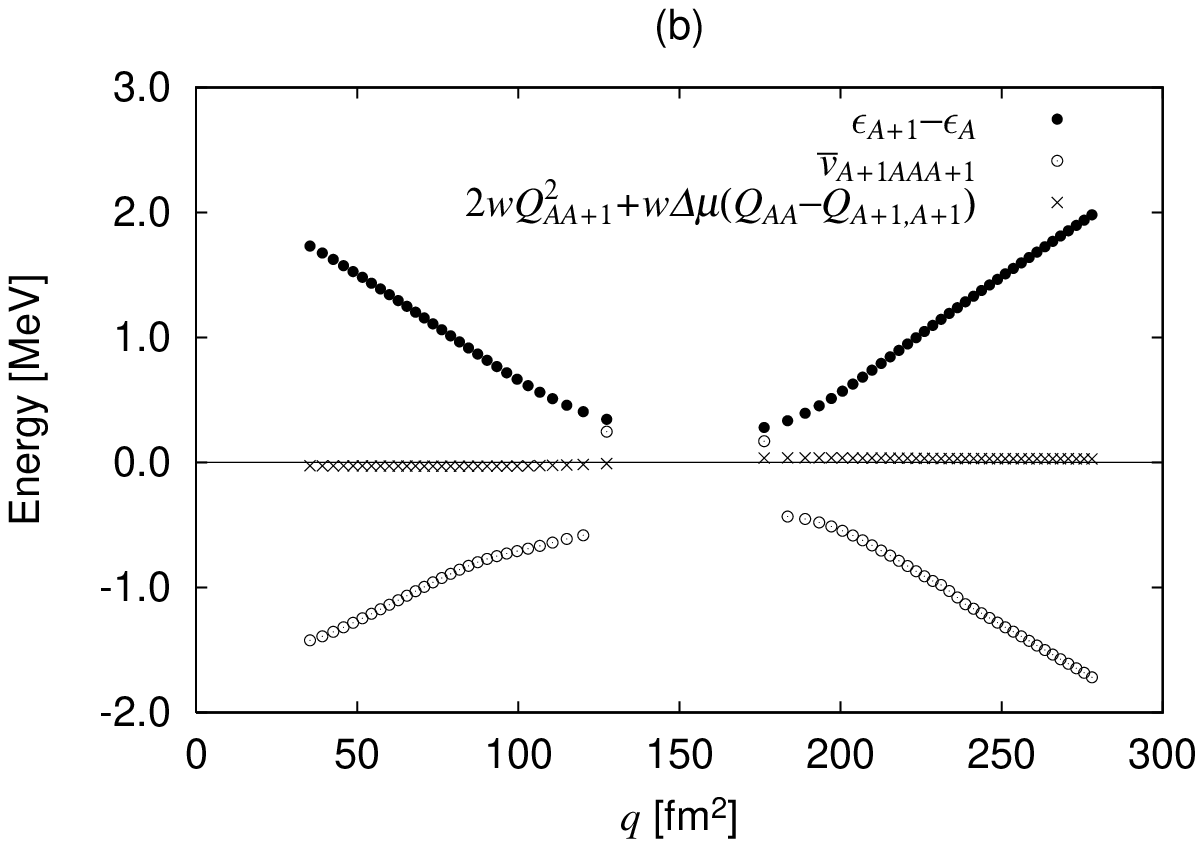}}
\caption{\label{Quan} (a) Energy difference between two specific orbits and quantum fluctuations as a function of quadrupole moment; (b) s.p. energy difference, two-body residual interaction and the sum of deformation fluctuation and quadrupole deformation.}
\end{figure}

\begin{figure}
\epsfxsize=8.0cm
\centerline{\epsffile{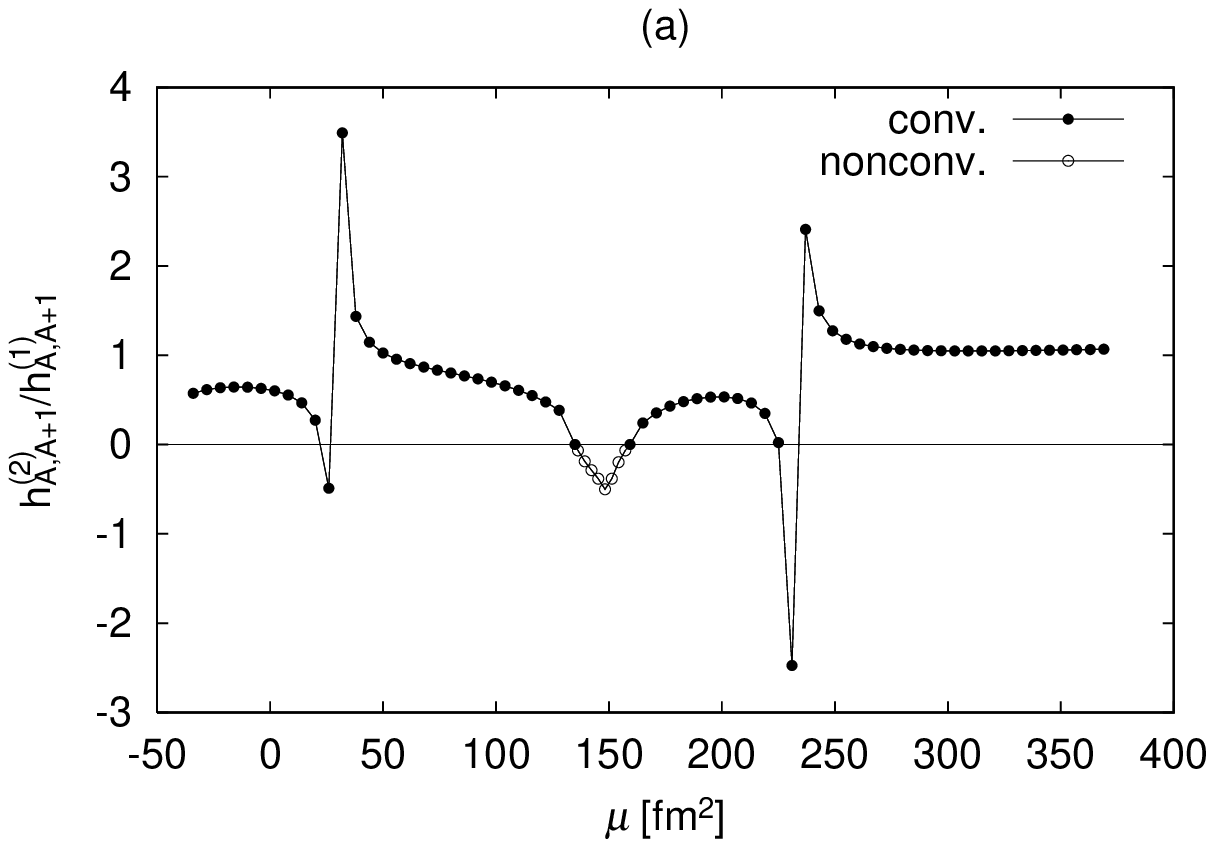}}
\centerline{\epsffile{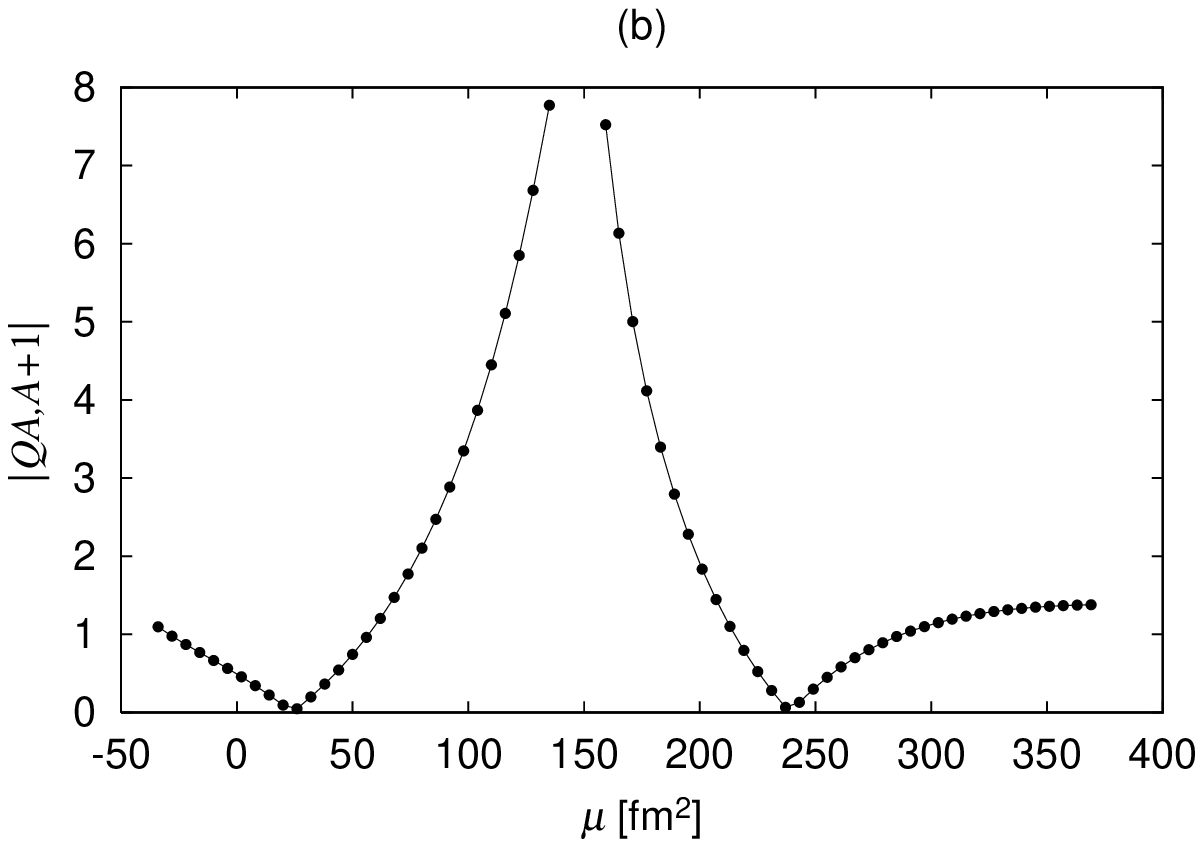}}
\caption{\label{Cond} (a) Ratio of off-diagonal components between the first and second iterations for each given $\mu$ state. Conv. and nonconv. stand for the convergent CHF states and nonconvergent region. (b) Absolute value of off-diagonal quadrupole moment as a function of quadrupole parameter $\mu$. }
\end{figure}

The physical meaning of the condition in Eq.~(\ref{con}) is clear: two-body correlation between nucleons can be successfully incorporated into the mean field as much as possible when the energy difference between two interacting orbits is not smaller than the quantum fluctuations coming from the two-body residual interaction and quadrupole deformation. 

Fig.~{\ref{Quan}}(a) gives the s.p. energy difference between two specific orbits $\varphi_A(q)$ and $\varphi_{A+1}(q)$, and the quantum fluctuations which are the sum of $\bar v_{A+1AAA+1}$, $2w{Q^2_{A,A+1}}$ and $w\Delta\mu(Q_{A,A}-Q_{A+1,A+1})$. 
In this figure and in Fig.~{\ref{Cond}}, the $q_0$-representation is used for any value of $q$ in the convergent region, and $\Delta q$ ($\Delta \mu$) is numerically determined by the configuration dictated method.
In the nonconvergent region, the $q_0$-representation is fixed at the critical point and $\Delta \mu$ is treated as a changeable parameter.
Effects coming from the higher-order term $O(b)$ is neglected. 
Our numerical results clearly show how nicely the analytic condition in Eq.~(\ref{con}) holds. 
Near the critical point, one may observe that the difference of s.p. energies is almost equal to the quantum fluctuations, i.e.,  $\epsilon_{A+1}(q)-\epsilon_{A}(q) \approx  \bar v_{A+1AAA+1}+2w{Q^2_{A,A+1}}+w\Delta\mu (Q_{A,A}-Q_{A+1,A+1})$. 
Figure~{\ref{Quan}}(b) shows the difference of s.p. energies $\epsilon_{A+1}(q)-\epsilon_{A}(q)$, the two-body residual interaction $ \bar v_{A+1AAA+1}$, and the effects from quadrupole deformation $2w{Q^2_{A,A+1}}+w\Delta\mu (Q_{A,A}-Q_{A+1,A+1})$, separately. One may observe that a competition between one-body potential $\epsilon_{A+1}-\epsilon_{A}$ and the two-body residual interaction $\bar v_{A+1AAA+1}$ indeed plays a dominant role in determining whether or not the concept of CHF mean field is realized.

Figure~{\ref{Cond}}(a) depicts the ratio $h^{(2)}_{A,A+1}/h^{(1)}_{A,A+1}$ as a function of quadrupole parameter $\mu$ (since the average quadrupole moment $q$ has no meaning in the nonconvergent region). One may see that in the convergent region the ratio is always positive except for some exceptional points, where the deformation fluctuation between two-specific orbits $Q_{A, A+1}$ is almost zero as shown in Fig.~{\ref{Cond}}(b). Near the critical point, the ratio is going to reach zero.  When one slightly decreases the deformation $\Delta\mu$ from the critical point, the ratio becomes negative where the CHF calculation meets a difficulty of nonconvergence. 
After the nonconvergent region, the ratio becomes positive, and the convergent state appears again.  
From these numerical calculation, it is clear that our condition in Eq.~(\ref{condition}) actually works well in evaluating whether the concept of the mean field is realized or not in the many-Fermion system. 

We should now comment on the exceptional points discussed above.
Since the $ph$ component of the quadrupole operator $|Q_{A, A+1}|$ is almost zero, which just corresponds to a full alignment state created by the angular momentum constraint in the cranked HF theory, a new deformed state can not be generated by the quadrupole operator within the truncated two dimensional subspace.
What actually happens at these exceptional points attributes to the importance of the other part of the s.p. space outside the two dimensional subspace spanned by $\varphi_A(q_0)$ and $\varphi_{A+1}(q_0)$.
Since our simple analytic understanding that uses the $2\times 2$ Hamilton matrix does not work in this specific situation, an appearance of exceptional points is out of our present discussion.

\section{summary}
The present work does not include the pairing correlation. Including the pairing correlation might make the situation so far discussed more complicated because of many dynamical competitions that exist not only between the $ph$-type two-body residual interaction and the HF potential, but also the $pp$-type two-body residual interaction and the pairing potential; their cross effects also must be considered. Moreover, the two mean fields characterized by different configurations are mixed up by the pairing correlation, and the $uv$ factor introduced by the BCS theory obscures the concept of the configuration.  Although some numerical evidence on an applicability of the CHFB theory near level crossing region has been discussed in Ref.~\cite{Guo3}, the role of pairing correlation on the level-crossing dynamics will be discussed in a separated paper.

In summary, the nonconvergent difficulty in level repulsive region is discussed analytically in the self-consistent mean field theory. It turns out that the CHF mean field  breaks down when the quantum fluctuations coming from two-body residual interaction and quadrupole deformation become larger than an energy difference between two avoided crossing orbits. 
Deriving the analytic condition, we make it clear that  the competition between one-body potential and quantum fluctuations mainly coming from two-body residual interaction plays an important role whether the self-consistent CHF mean field is realized or not.
However, the breakdown of the CHF mean field at certain range of the quadrupole deformation $q$ does not mean that introducing some quantum mechanical treatment is necessary, because the constrained operator $\hat Q$ in the CHF theory is put in by hand.
That is, the system does not like to be elongated nor contracted along a given direction of quadrupole deformation any more, but it likes to develop toward a direction chosen by itself.
Further microscopic investigation is needed to answer the very interesting conclusion in Ref.~\cite{RB89} by introducing {\it dynamical} constrained operators based on the self-consistent collective coordinate (SCC) method \cite{SCC1, SCC2}, because a diabolic point related to the level crossing indicates an existence of missing degree of freedom.

\begin{acknowledgments}
This work is supported in part by the Japan Society for the Promotion of Science (JSPS) and the China National Natural Science Foundation (CNSF) as the bilateral program between Japan and China. E. G. Zhao acknowledges the support by Natural Science Foundation of China under Grant No. 10375001 and 10435010,  the China  Major Basic Research Development Program under Grant No. G2000-0774-07, the Knowledge Innovation Project of the Chinese Academy of Sciences under Grant No. KJCX2-SW-N02.
\end{acknowledgments}

\bibliography{new}

\begin{thebibliography}{33}
\expandafter\ifx\csname natexlab\endcsname\relax\def\natexlab#1{#1}\fi
\expandafter\ifx\csname bibnamefont\endcsname\relax
  \def\bibnamefont#1{#1}\fi
\expandafter\ifx\csname bibfnamefont\endcsname\relax
  \def\bibfnamefont#1{#1}\fi
\expandafter\ifx\csname citenamefont\endcsname\relax
  \def\citenamefont#1{#1}\fi
\expandafter\ifx\csname url\endcsname\relax
  \def\url#1{\texttt{#1}}\fi
\expandafter\ifx\csname urlprefix\endcsname\relax\def\urlprefix{URL }\fi
\providecommand{\bibinfo}[2]{#2}
\providecommand{\eprint}[2][]{\url{#2}}

\bibitem[{\citenamefont{Arias et~al.}(2003)\citenamefont{Arias, Dukelsky, and
  Garcia-Ramos}}]{JMA03}
\bibinfo{author}{\bibfnamefont{J.~M.} \bibnamefont{Arias}},
  \bibinfo{author}{\bibfnamefont{J.}~\bibnamefont{Dukelsky}}, \bibnamefont{and}
  \bibinfo{author}{\bibfnamefont{J.~E.} \bibnamefont{Garcia-Ramos}},
  \bibinfo{journal}{Phys. Rev. Lett} \textbf{\bibinfo{volume}{91}},
  \bibinfo{pages}{162502} (\bibinfo{year}{2003}).

\bibitem[{\citenamefont{Gonz{\'a}lez-F{\'e}rez and Dehesa}(2003)}]{RGF03}
\bibinfo{author}{\bibfnamefont{R.}~\bibnamefont{Gonz{\'a}lez-F{\'e}rez}}
  \bibnamefont{and} \bibinfo{author}{\bibfnamefont{J.~S.}
  \bibnamefont{Dehesa}}, \bibinfo{journal}{Phys. Rev. Lett}
  \textbf{\bibinfo{volume}{91}}, \bibinfo{pages}{113001}
  (\bibinfo{year}{2003}).

\bibitem[{\citenamefont{Intonti et~al.}(2001)\citenamefont{Intonti, Emiliani,
  Lienau, Elsaesser, Savona, Runge, Zimmermann, Notzel, and Ploog}}]{FI01}
\bibinfo{author}{\bibfnamefont{F.}~\bibnamefont{Intonti}},
  \bibinfo{author}{\bibfnamefont{V.}~\bibnamefont{Emiliani}},
  \bibinfo{author}{\bibfnamefont{C.}~\bibnamefont{Lienau}},
  \bibinfo{author}{\bibfnamefont{T.}~\bibnamefont{Elsaesser}},
  \bibinfo{author}{\bibfnamefont{V.}~\bibnamefont{Savona}},
  \bibinfo{author}{\bibfnamefont{E.}~\bibnamefont{Runge}},
  \bibinfo{author}{\bibfnamefont{R.}~\bibnamefont{Zimmermann}},
  \bibinfo{author}{\bibfnamefont{R.}~\bibnamefont{Notzel}}, \bibnamefont{and}
  \bibinfo{author}{\bibfnamefont{K.~H.} \bibnamefont{Ploog}},
  \bibinfo{journal}{Phys. Rev. Lett} \textbf{\bibinfo{volume}{87}},
  \bibinfo{pages}{076801} (\bibinfo{year}{2001}).

\bibitem[{\citenamefont{Dembowski et~al.}(2001)\citenamefont{Dembowski, Graf,
  Harney, Heine, Heiss, Rehfeld, and Richter}}]{CD01}
\bibinfo{author}{\bibfnamefont{C.}~\bibnamefont{Dembowski}},
  \bibinfo{author}{\bibfnamefont{H.~D.} \bibnamefont{Graf}},
  \bibinfo{author}{\bibfnamefont{H.~L.} \bibnamefont{Harney}},
  \bibinfo{author}{\bibfnamefont{A.}~\bibnamefont{Heine}},
  \bibinfo{author}{\bibfnamefont{W.~D.} \bibnamefont{Heiss}},
  \bibinfo{author}{\bibfnamefont{H.}~\bibnamefont{Rehfeld}}, \bibnamefont{and}
  \bibinfo{author}{\bibfnamefont{A.}~\bibnamefont{Richter}},
  \bibinfo{journal}{Phys. Rev. Lett} \textbf{\bibinfo{volume}{86}},
  \bibinfo{pages}{787} (\bibinfo{year}{2001}).

\bibitem[{\citenamefont{Hamamoto}(1976)}]{Hama76}
\bibinfo{author}{\bibfnamefont{I.}~\bibnamefont{Hamamoto}},
  \bibinfo{journal}{Nucl. Phys. A} \textbf{\bibinfo{volume}{271}},
  \bibinfo{pages}{15} (\bibinfo{year}{1976}).

\bibitem[{\citenamefont{Strutinsky}(1977)}]{Stru77}
\bibinfo{author}{\bibfnamefont{V.~M.} \bibnamefont{Strutinsky}},
  \bibinfo{journal}{Z. Phys. A} \textbf{\bibinfo{volume}{280}},
  \bibinfo{pages}{99} (\bibinfo{year}{1977}).

\bibitem[{\citenamefont{Nazarewicz}(1993)}]{Naz93}
\bibinfo{author}{\bibfnamefont{W.}~\bibnamefont{Nazarewicz}},
  \bibinfo{journal}{Nucl. Phys. A} \textbf{\bibinfo{volume}{557}},
  \bibinfo{pages}{489c} (\bibinfo{year}{1993}).

\bibitem[{\citenamefont{Dobaczewski and Dudek}(2000)}]{Dob00}
\bibinfo{author}{\bibfnamefont{J.}~\bibnamefont{Dobaczewski}} \bibnamefont{and}
  \bibinfo{author}{\bibfnamefont{J.}~\bibnamefont{Dudek}},
  \bibinfo{journal}{Comput. Phys. Commun} \textbf{\bibinfo{volume}{131}},
  \bibinfo{pages}{164} (\bibinfo{year}{2000}).

\bibitem[{\citenamefont{Bengtsson and Nazarewicz}(1989)}]{RB89}
\bibinfo{author}{\bibfnamefont{R.}~\bibnamefont{Bengtsson}} \bibnamefont{and}
  \bibinfo{author}{\bibfnamefont{W.}~\bibnamefont{Nazarewicz}},
  \bibinfo{journal}{Z. Phys. A} \textbf{\bibinfo{volume}{334}},
  \bibinfo{pages}{269} (\bibinfo{year}{1989}).

\bibitem[{\citenamefont{Bengtsson and Ragnarsson}(1985)}]{TB}
\bibinfo{author}{\bibfnamefont{T.}~\bibnamefont{Bengtsson}} \bibnamefont{and}
  \bibinfo{author}{\bibfnamefont{I.}~\bibnamefont{Ragnarsson}},
  \bibinfo{journal}{Nucl. \ Phys. A} \textbf{\bibinfo{volume}{436}},
  \bibinfo{pages}{14} (\bibinfo{year}{1985}).

\bibitem[{\citenamefont{Bengtsson et~al.}(1994)\citenamefont{Bengtsson,
  Bengtsson, Bergstr{\"o}m, Ryde, and Hagemann}}]{RB}
\bibinfo{author}{\bibfnamefont{R.}~\bibnamefont{Bengtsson}},
  \bibinfo{author}{\bibfnamefont{T.}~\bibnamefont{Bengtsson}},
  \bibinfo{author}{\bibfnamefont{M.}~\bibnamefont{Bergstr{\"o}m}},
  \bibinfo{author}{\bibfnamefont{H.}~\bibnamefont{Ryde}}, \bibnamefont{and}
  \bibinfo{author}{\bibfnamefont{G.~B.} \bibnamefont{Hagemann}},
  \bibinfo{journal}{Nucl. \ Phys. A} \textbf{\bibinfo{volume}{569}},
  \bibinfo{pages}{469} (\bibinfo{year}{1994}).

\bibitem[{\citenamefont{Bengtsson}(1989)}]{TB89}
\bibinfo{author}{\bibfnamefont{T.}~\bibnamefont{Bengtsson}},
  \bibinfo{journal}{Nucl. \ Phys. A} \textbf{\bibinfo{volume}{496}},
  \bibinfo{pages}{56} (\bibinfo{year}{1989}).

\bibitem[{\citenamefont{Axelsson
  et~al.}(2002{\natexlab{a}})\citenamefont{Axelsson, Bengtsson, and
  Nyberg}}]{AA021}
\bibinfo{author}{\bibfnamefont{A.}~\bibnamefont{Axelsson}},
  \bibinfo{author}{\bibfnamefont{R.}~\bibnamefont{Bengtsson}},
  \bibnamefont{and} \bibinfo{author}{\bibfnamefont{J.}~\bibnamefont{Nyberg}},
  \bibinfo{journal}{Nucl. Phys. A} \textbf{\bibinfo{volume}{708}},
  \bibinfo{pages}{226} (\bibinfo{year}{2002}{\natexlab{a}}).

\bibitem[{\citenamefont{Axelsson
  et~al.}(2002{\natexlab{b}})\citenamefont{Axelsson, Bengtsson, and
  Nyberg}}]{AA022}
\bibinfo{author}{\bibfnamefont{A.}~\bibnamefont{Axelsson}},
  \bibinfo{author}{\bibfnamefont{R.}~\bibnamefont{Bengtsson}},
  \bibnamefont{and} \bibinfo{author}{\bibfnamefont{J.}~\bibnamefont{Nyberg}},
  \bibinfo{journal}{Nucl. Phys. A} \textbf{\bibinfo{volume}{708}},
  \bibinfo{pages}{245} (\bibinfo{year}{2002}{\natexlab{b}}).

\bibitem[{\citenamefont{Iwasawa et~al.}(1994)\citenamefont{Iwasawa, Sakata,
  Hashimoto, and Terasaki}}]{CD1}
\bibinfo{author}{\bibfnamefont{K.}~\bibnamefont{Iwasawa}},
  \bibinfo{author}{\bibfnamefont{F.}~\bibnamefont{Sakata}},
  \bibinfo{author}{\bibfnamefont{Y.}~\bibnamefont{Hashimoto}},
  \bibnamefont{and} \bibinfo{author}{\bibfnamefont{J.}~\bibnamefont{Terasaki}},
  \bibinfo{journal}{Prog. Theor. Phys.} \textbf{\bibinfo{volume}{92}},
  \bibinfo{pages}{1119} (\bibinfo{year}{1994}).

\bibitem[{\citenamefont{Guo et~al.}(2004{\natexlab{a}})\citenamefont{Guo,
  Sakata, and Zhao}}]{Guo3}
\bibinfo{author}{\bibfnamefont{L.}~\bibnamefont{Guo}},
  \bibinfo{author}{\bibfnamefont{F.}~\bibnamefont{Sakata}}, \bibnamefont{and}
  \bibinfo{author}{\bibfnamefont{E.-G.} \bibnamefont{Zhao}},
  \bibinfo{journal}{Nucl. Phys. A} \textbf{\bibinfo{volume}{740}},
  \bibinfo{pages}{59} (\bibinfo{year}{2004}{\natexlab{a}}).

\bibitem[{\citenamefont{Guo et~al.}(2004{\natexlab{b}})\citenamefont{Guo,
  Sakata, and Zhao}}]{Guo2}
\bibinfo{author}{\bibfnamefont{L.}~\bibnamefont{Guo}},
  \bibinfo{author}{\bibfnamefont{F.}~\bibnamefont{Sakata}}, \bibnamefont{and}
  \bibinfo{author}{\bibfnamefont{E.-G.} \bibnamefont{Zhao}},
  \bibinfo{journal}{nucl-th/0407031}  (\bibinfo{year}{2004}{\natexlab{b}}).

\bibitem[{\citenamefont{Gogny}(1973)}]{Gog1}
\bibinfo{author}{\bibfnamefont{D.}~\bibnamefont{Gogny}}, in
  \emph{\bibinfo{booktitle}{Nuclear Self-Consistent Fields}}, edited by
  \bibinfo{editor}{\bibfnamefont{G.}~\bibnamefont{Ripka}} \bibnamefont{and}
  \bibinfo{editor}{\bibfnamefont{M.}~\bibnamefont{Porneuf}}
  (\bibinfo{address}{North-Holland, Amsterdam}, \bibinfo{year}{1973}), p.
  \bibinfo{pages}{333}.

\bibitem[{\citenamefont{Decharg\'{e} and Gogny}(1980)}]{Gog2}
\bibinfo{author}{\bibfnamefont{J.}~\bibnamefont{Decharg\'{e}}}
  \bibnamefont{and} \bibinfo{author}{\bibfnamefont{D.}~\bibnamefont{Gogny}},
  \bibinfo{journal}{Phys. \ Rev. C} \textbf{\bibinfo{volume}{21}},
  \bibinfo{pages}{1568} (\bibinfo{year}{1980}).

\bibitem[{\citenamefont{Girod and Grammaticos}(1983)}]{Gog3}
\bibinfo{author}{\bibfnamefont{M.}~\bibnamefont{Girod}} \bibnamefont{and}
  \bibinfo{author}{\bibfnamefont{B.}~\bibnamefont{Grammaticos}},
  \bibinfo{journal}{Phys. \ Rev. C} \textbf{\bibinfo{volume}{27}},
  \bibinfo{pages}{2317} (\bibinfo{year}{1983}).

\bibitem[{\citenamefont{Berger et~al.}(1984)\citenamefont{Berger, Girod, and
  Gogny}}]{Gog4}
\bibinfo{author}{\bibfnamefont{J.~F.} \bibnamefont{Berger}},
  \bibinfo{author}{\bibfnamefont{M.}~\bibnamefont{Girod}}, \bibnamefont{and}
  \bibinfo{author}{\bibfnamefont{D.}~\bibnamefont{Gogny}},
  \bibinfo{journal}{Nucl. \ Phys. A} \textbf{\bibinfo{volume}{428}},
  \bibinfo{pages}{23c} (\bibinfo{year}{1984}).

\bibitem[{\citenamefont{Berger et~al.}(1989)\citenamefont{Berger, Girod, and
  Gogny}}]{Gog5}
\bibinfo{author}{\bibfnamefont{J.~F.} \bibnamefont{Berger}},
  \bibinfo{author}{\bibfnamefont{M.}~\bibnamefont{Girod}}, \bibnamefont{and}
  \bibinfo{author}{\bibfnamefont{D.}~\bibnamefont{Gogny}},
  \bibinfo{journal}{Nucl. \ Phys. A} \textbf{\bibinfo{volume}{502}},
  \bibinfo{pages}{85c} (\bibinfo{year}{1989}).

\bibitem[{\citenamefont{Berger et~al.}(1991)\citenamefont{Berger, Girod, and
  Gogny}}]{Gog6}
\bibinfo{author}{\bibfnamefont{J.~F.} \bibnamefont{Berger}},
  \bibinfo{author}{\bibfnamefont{M.}~\bibnamefont{Girod}}, \bibnamefont{and}
  \bibinfo{author}{\bibfnamefont{D.}~\bibnamefont{Gogny}},
  \bibinfo{journal}{Comput. Phys. Commun.} \textbf{\bibinfo{volume}{63}},
  \bibinfo{pages}{365} (\bibinfo{year}{1991}).

\bibitem[{\citenamefont{Flocard et~al.}(1973)\citenamefont{Flocard, Quentin,
  Kerman, and Vautherin}}]{HFL}
\bibinfo{author}{\bibfnamefont{H.}~\bibnamefont{Flocard}},
  \bibinfo{author}{\bibfnamefont{P.}~\bibnamefont{Quentin}},
  \bibinfo{author}{\bibfnamefont{A.~K.} \bibnamefont{Kerman}},
  \bibnamefont{and}
  \bibinfo{author}{\bibfnamefont{D.}~\bibnamefont{Vautherin}},
  \bibinfo{journal}{Nucl. Phys. A} \textbf{\bibinfo{volume}{203}},
  \bibinfo{pages}{433} (\bibinfo{year}{1973}).

\bibitem[{\citenamefont{Dobaczewski
  et~al.}(2000{\natexlab{a}})\citenamefont{Dobaczewski, Dudek, Rohozinski, and
  Werner}}]{JD1}
\bibinfo{author}{\bibfnamefont{J.}~\bibnamefont{Dobaczewski}},
  \bibinfo{author}{\bibfnamefont{J.}~\bibnamefont{Dudek}},
  \bibinfo{author}{\bibfnamefont{S.~G.} \bibnamefont{Rohozinski}},
  \bibnamefont{and} \bibinfo{author}{\bibfnamefont{T.~R.}
  \bibnamefont{Werner}}, \bibinfo{journal}{Phys. \ Rev. C}
  \textbf{\bibinfo{volume}{62}}, \bibinfo{pages}{014310}
  (\bibinfo{year}{2000}{\natexlab{a}}).

\bibitem[{\citenamefont{Dobaczewski
  et~al.}(2000{\natexlab{b}})\citenamefont{Dobaczewski, Dudek, Rohozinski, and
  Werner}}]{JD2}
\bibinfo{author}{\bibfnamefont{J.}~\bibnamefont{Dobaczewski}},
  \bibinfo{author}{\bibfnamefont{J.}~\bibnamefont{Dudek}},
  \bibinfo{author}{\bibfnamefont{S.~G.} \bibnamefont{Rohozinski}},
  \bibnamefont{and} \bibinfo{author}{\bibfnamefont{T.~R.}
  \bibnamefont{Werner}}, \bibinfo{journal}{Phys. \ Rev. C}
  \textbf{\bibinfo{volume}{62}}, \bibinfo{pages}{014311}
  (\bibinfo{year}{2000}{\natexlab{b}}).

\bibitem[{\citenamefont{Dobaczewski and Dudek}(1995)}]{ref1}
\bibinfo{author}{\bibfnamefont{J.}~\bibnamefont{Dobaczewski}} \bibnamefont{and}
  \bibinfo{author}{\bibfnamefont{J.}~\bibnamefont{Dudek}},
  \bibinfo{journal}{Phys. Rev. C} \textbf{\bibinfo{volume}{52}},
  \bibinfo{pages}{1827} (\bibinfo{year}{1995}).

\bibitem[{\citenamefont{Molique et~al.}(2000)\citenamefont{Molique,
  Dobaczewski, and Dudek}}]{ref2}
\bibinfo{author}{\bibfnamefont{H.}~\bibnamefont{Molique}},
  \bibinfo{author}{\bibfnamefont{J.}~\bibnamefont{Dobaczewski}},
  \bibnamefont{and} \bibinfo{author}{\bibfnamefont{J.}~\bibnamefont{Dudek}},
  \bibinfo{journal}{Phys. Rev. C} \textbf{\bibinfo{volume}{61}},
  \bibinfo{pages}{044304} (\bibinfo{year}{2000}).

\bibitem[{\citenamefont{Satula et~al.}(1996)\citenamefont{Satula, Dobaczewski,
  Dudek, and Nazarewicz}}]{ref3}
\bibinfo{author}{\bibfnamefont{W.}~\bibnamefont{Satula}},
  \bibinfo{author}{\bibfnamefont{J.}~\bibnamefont{Dobaczewski}},
  \bibinfo{author}{\bibfnamefont{J.}~\bibnamefont{Dudek}}, \bibnamefont{and}
  \bibinfo{author}{\bibfnamefont{W.}~\bibnamefont{Nazarewicz}},
  \bibinfo{journal}{Phys. Rev. Lett} \textbf{\bibinfo{volume}{77}},
  \bibinfo{pages}{5182} (\bibinfo{year}{1996}).

\bibitem[{\citenamefont{Hackman et~al.}(1995)\citenamefont{Hackman, Wadsworth,
  Haslip, Clark, Dobaczewski, Dudek, Flibotte, and {\it et al.}}}]{ref4}
\bibinfo{author}{\bibfnamefont{G.}~\bibnamefont{Hackman}},
  \bibinfo{author}{\bibfnamefont{R.}~\bibnamefont{Wadsworth}},
  \bibinfo{author}{\bibfnamefont{D.~S.} \bibnamefont{Haslip}},
  \bibinfo{author}{\bibfnamefont{R.~M.} \bibnamefont{Clark}},
  \bibinfo{author}{\bibfnamefont{J.}~\bibnamefont{Dobaczewski}},
  \bibinfo{author}{\bibfnamefont{J.}~\bibnamefont{Dudek}},
  \bibinfo{author}{\bibfnamefont{S.}~\bibnamefont{Flibotte}}, \bibnamefont{and}
  \bibinfo{author}{\bibfnamefont{K.~H.} \bibnamefont{{\it et al.}}},
  \bibinfo{journal}{Phys. Rev. C} \textbf{\bibinfo{volume}{52}},
  \bibinfo{pages}{R2293} (\bibinfo{year}{1995}).

\bibitem[{\citenamefont{Audi et~al.}(1997)\citenamefont{Audi, Bersillon,
  Blachot, and Wapstra}}]{Audi}
\bibinfo{author}{\bibfnamefont{G.}~\bibnamefont{Audi}},
  \bibinfo{author}{\bibfnamefont{O.}~\bibnamefont{Bersillon}},
  \bibinfo{author}{\bibfnamefont{J.}~\bibnamefont{Blachot}}, \bibnamefont{and}
  \bibinfo{author}{\bibfnamefont{A.~H.} \bibnamefont{Wapstra}},
  \bibinfo{journal}{Nucl. Phys. A} \textbf{\bibinfo{volume}{624}},
  \bibinfo{pages}{1} (\bibinfo{year}{1997}).

\bibitem[{\citenamefont{Marumori et~al.}(1980)\citenamefont{Marumori, Maskawa,
  Sakata, and Kuriyama}}]{SCC1}
\bibinfo{author}{\bibfnamefont{T.}~\bibnamefont{Marumori}},
  \bibinfo{author}{\bibfnamefont{T.}~\bibnamefont{Maskawa}},
  \bibinfo{author}{\bibfnamefont{F.}~\bibnamefont{Sakata}}, \bibnamefont{and}
  \bibinfo{author}{\bibfnamefont{A.}~\bibnamefont{Kuriyama}},
  \bibinfo{journal}{Prog. Theor. Phys} \textbf{\bibinfo{volume}{64}},
  \bibinfo{pages}{1294} (\bibinfo{year}{1980}).

\bibitem[{\citenamefont{Sakata et~al.}(2001)\citenamefont{Sakata, Marumori,
  Hashimoto, and Yan}}]{SCC2}
\bibinfo{author}{\bibfnamefont{F.}~\bibnamefont{Sakata}},
  \bibinfo{author}{\bibfnamefont{T.}~\bibnamefont{Marumori}},
  \bibinfo{author}{\bibfnamefont{Y.}~\bibnamefont{Hashimoto}},
  \bibnamefont{and} \bibinfo{author}{\bibfnamefont{S.~W.} \bibnamefont{Yan}},
  \bibinfo{journal}{Suppl. Prog. Theor. Phys} \textbf{\bibinfo{volume}{141}},
  \bibinfo{pages}{1} (\bibinfo{year}{2001}).

\end{thebibliography}
\end{document}